\documentclass{article}
\usepackage[pdftex]{graphicx}

\usepackage{bm}
\usepackage{amsmath, amsthm, amssymb}
\usepackage{rotating}

\newcommand{\citeyear}{\cite}
\newcommand{\opencite}{\cite}
\newcommand{\shortcite}{\cite}
\newcommand{\degree}{\ensuremath{^\circ}}
\newcommand\pd[3][]{\frac{\partial^{#1} #2}{\partial{#3}^{#1}}}
{{}}
\newcommand{\ssu}{s_0}
\newcommand{\ssv}{s}
\newcommand{\lba}{\textsf{\textbf{a}}\quad}
\newcommand{\lbb}{\textsf{\textbf{b}}\quad}
\newcommand{\lbc}{\textsf{\textbf{c}}\quad}
\newcommand{\lbd}{\textsf{\textbf{d}}\quad}

\begin{document}

%

\title{Mathematical embryology: the fluid mechanics of nodal cilia$^*$}
\author{D.J. Smith$^{**1,2,3,4}$, A.A. Smith$^{1,2}$ \& J.R. Blake$^{1,2}$ \\
\small{$^1$School of Mathematics, University of Birmingham, Edgbaston,} \\
\small{Birmingham B15 2TT, United Kingdom} \\
\small{$^2$Centre for Human Reproductive Science, Birmingham Women's NHS Foundation Trust, Metchley Park Road,}\\ 
\small{Edgbaston, Birmingham B15 2TG, United Kingdom}\\
\small{$^3$School of Engineering \& $^4$Centre for Scientific Computing, University of Warwick,}\\
\small{Coventry CV4 7AL, United Kingdom}\\
\small{$^*$Preprint of an article to appear in:}\\
\small{Tuck Memorial Issue of the Journal of Engineering Mathematics, 2011}\\
\small{$^{**}$Corresponding author, email address: d.j.smith.2@bham.ac.uk} 
} 

\date{}

\maketitle

\begin{abstract}
Left-right symmetry breaking is critical to vertebrate embryonic development; in many species this process begins with cilia-driven flow in a structure termed the `node'. 
Primary `whirling' cilia, tilted towards the posterior, transport morphogen-containing vesicles towards the left, initiating left-right asymmetric development. We review recent theoretical models based on the point-force stokeslet and point-torque rotlet singularities, explaining how rotation and surface-tilt produce directional flow. 
Analysis of image singularity systems enforcing the no-slip condition shows how tilted rotation produces a far-field `stresslet' directional flow, 
and how time-dependent point-force and time-independent point-torque models are in this respect equivalent.  Associated slender body theory analysis is reviewed; this approach enables efficient and accurate simulation of three-dimensional time-dependent flow, time-dependence being essential in predicting features of the flow such as chaotic advection, which have subsequently been determined experimentally. A new model for the nodal flow utilising the regularized stokeslet method is developed, to model the effect of the overlying Reichert's membrane. Velocity fields and particle paths within the enclosed domain are computed and compared with the flow profiles predicted by previous `membrane-less' models. Computations confirm that the presence of the membrane produces flow-reversal in the upper region, but no continuous region of reverse flow close to the epithelium. The stresslet far-field is no longer evident in the membrane model, due to the depth of the cavity being of similar magnitude to the cilium length.
Simulations predict that vesicles released within one cilium length of the epithelium are generally transported to the left via a `loopy drift' motion, sometimes involving highly unpredictable detours around leftward cilia. Particles released just above the cilia tips were not predicted to reach to the extreme edges of the node, but rather are returned to the right by the counterflow. Flow to the right and left of the cilia array is of very small magnitude, suggesting that effective transport of particles to the extremities of the node requires cilia to be distributed all the way to the edges.
There is no continuous layer of rightward flow close to the epithelium, except for a region close to the posterior edge of the node. Future work will involve investigating issues such as the precise shape of the node and cilia distribution and the effect of advection and diffusion on morphogens, hence explaining more fully the role of fluid mechanics in this vital developmental process.
\end{abstract}


\section{Personal reflection}

It is a great honour to be invited to contribute to the volume in memory of Professor Ernie Tuck, the last authors' undergraduate project supervisor, mentor and friend for life, his influence being evident in this paper. During the period 1968--69 Tuck introduced the last author to slender body theory in Stokes flow, based on his earlier papers \shortcite{Tuck64,Tuck68a}. Part of his project work has recently been published, $40$ years on \shortcite{Blake10}. The application of slender body theory to micro-organism locomotion led to studying ciliary propulsion, the study of which began during Tuck's period at Caltech under the direction of the distinguished hydrodynamicist, T.Y. Wu.  While at Caltech, Tuck \shortcite{Tuck68b} extended the famous Taylor \shortcite{Taylor51} analysis of a swimming sheet in a viscous fluid by adding inertia and tangential motion to the analysis. As a result Tuck introduced the last author to the paper by Lighthill \shortcite{Lighthill52} on the squirming motion of a sphere in a viscous liquid, which was found to contain an analytical error. This later led to the last author undertaking a Ph.D under the supervision of Sir James Lighthill, the recently appointed Lucasian Professor of Mathematics at the University of Cambridge. One of the first actions, encouraged by Lighthill, was to publish a corrected version of the paper \cite{Blake71b} but with an application directed at ciliary propulsion. Lighthill \shortcite{Lighthill73} reported on the outcome of some of these studies as one of the invited speakers at the ICTAM Congress in Moscow in $1972$. Thirty six years later, in $2008$, Ernie Tuck was the president of the organising committee of the highly successful ICTAM Congress held in Adelaide. This paper will incorporate the ideas originally initiated during this $1968$--$69$ period in the study of one of the most exciting current areas of science, embryonic development, an area which has previously attracted the attention of such luminaries as Alan Turing \shortcite{Turing52} and Francis Crick \shortcite{Crick70}.

\section{Introduction}\label{section:introduction}
We begin by reviewing the relevant biology of nodal cilia and left-right symmetry-breaking, and briefly outlining experimental and theoretical research into this process.
In subsequent sections of the paper we review current knowledge of the fluid mechanics in more detail, focusing on the use of singularities in explaining and modelling the flow. We then develop a new model which takes into account the effect of an overlying membrane, and present computational calculations showing its effect on the fluid flow and particle transport.

On first impressions, the bodies of mammals are essentially bilaterally symmetric. Beneath the surface it is very different however, with symmetry being broken in an organised way, the human heart for example being on the individual's left, the liver on the right (figure 1a). The conventional left-right axis designation in this figure, with the `left' of the individual appearing on the right of the figure will be used throughout this paper. This symmetry and asymmetry is laid down in the early stages of development, and is critical to embryonic survival. In this section we shall briefly describe the relevant biology, before addressing the fluid mechanics; for detailed review, see Hirokawa et al. \shortcite{Hirokawa09}. 

\begin{figure}
$
\begin{array}{c}
  \begin{array}{cc}
   {
    \includegraphics[scale=0.40]{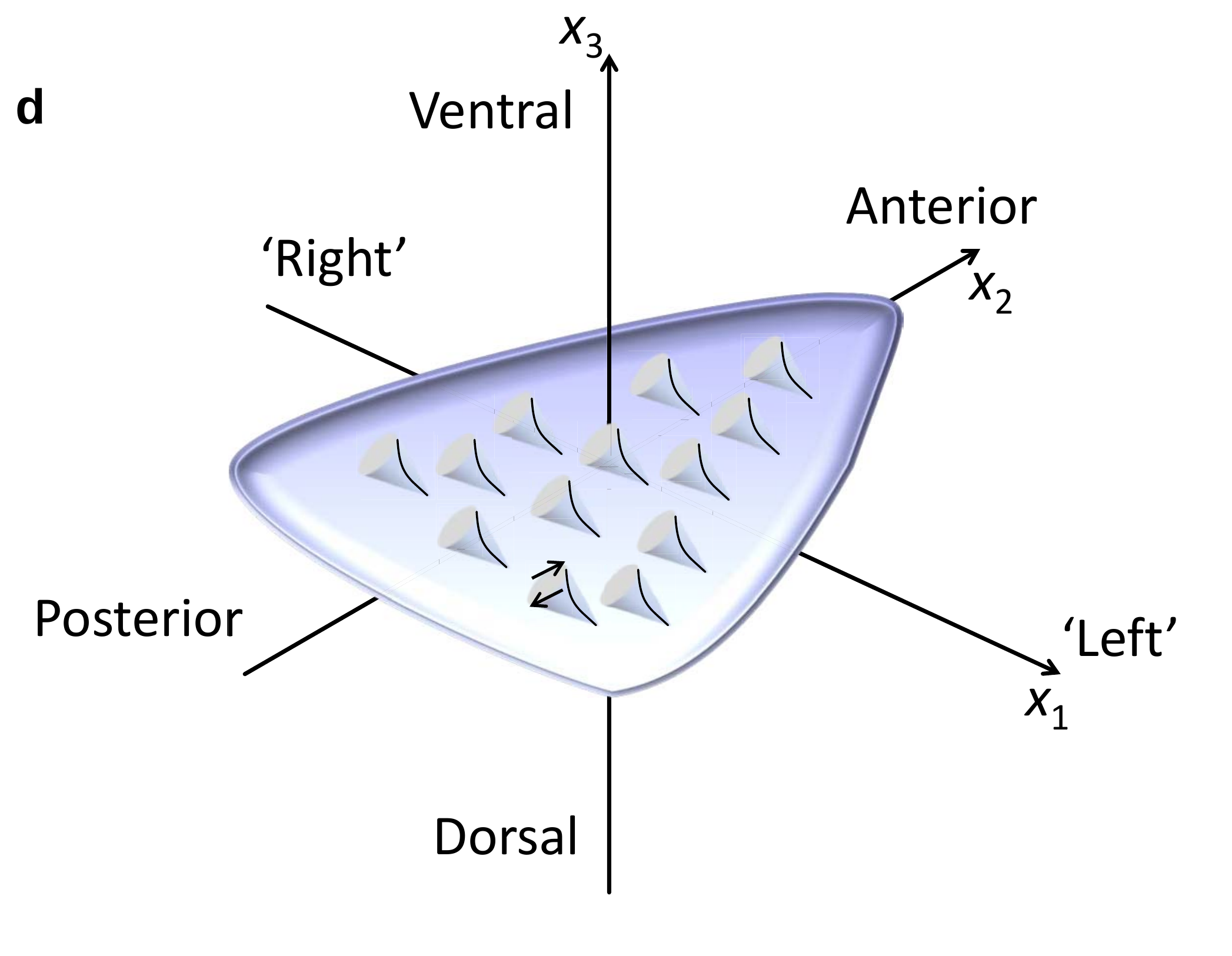}
   }
 &         
  \end{array}\\
 \begin{array}{cc}
   \raisebox{4mm}{ 
     \includegraphics[scale=0.56]{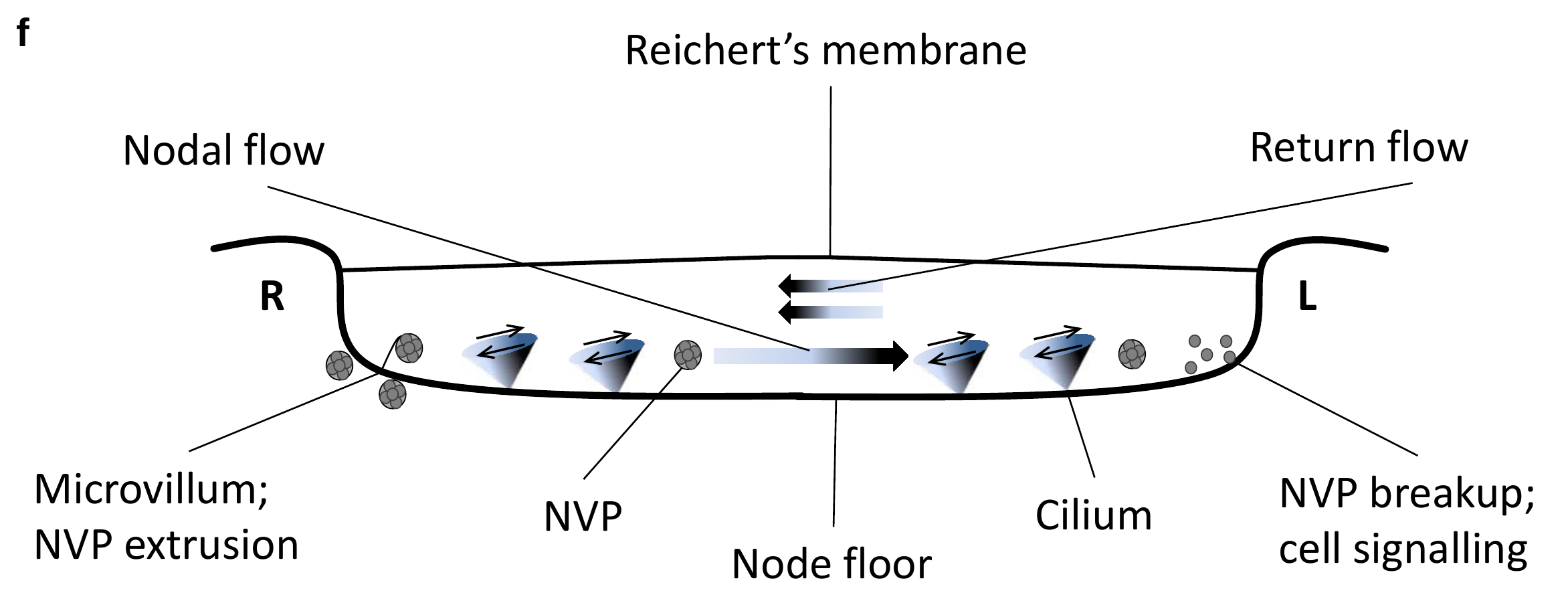}
    } &
  \end{array}
\\
\end{array}
$
\caption{The role of cilia in generating the mammalian body plan. (a) Normal organ placement and (b) mirrored organ placement in situs inversus (copyright figure not included in arxiv version). (c) The ventral side of the mouse embryo at eight days post-fertilisation, with Reichert's membrane removed, showing the anterior-posterior and left-right axes, the ventral node being the depression structure highlighted (copyright figure not included in arxiv version). Abbreviations: VN, ventral node; NP, notochordal plate; FG, foregut. (d) Schematic of the ventral node, showing how the biological axis designations correspond to the Cartesian axes used in the mathematical model of this paper. The node is shown looking down onto the ventral surface of the embryo; the cones represent the approximate beat pattern of the cilia. (e) Electron micrograph of the nodal floor and nodal cilia (copyright figure not included in arxiv version). (f) Schematic section of the node cavity, viewed from the posterior, looking towards the anterior, with the `left' of the embryo being conventionally shown on the right of the diagram. The `leftward' mainstream flow is balanced by a rightward return flow induced by Reichert's membrane; the existence of a rightward return flow close to the node floor is a subject of continuing discussion \shortcite{Cartwright08}.}
\label{fig1}
\end{figure}

The breaking of the left-right axis occurs after the establishment of the anterior-posterior and dorsal-ventral axes, but until the mid 1990s there was little understanding of the mechanisms involved. 
Kartagener in the 1930s, and others previously (for historical review, see \opencite{Berdon04a,Berdon04}), had identified a triad syndrome of respiratory problems, male infertility and situs inversus, the latter being the lateral transposition of the internal organs (figure 1b). The common factor causing these symptoms was not however understood.
A serendipitous discovery occurred in 1974 when Afzelius examined the doctoral thesis of H. Pedersen. A link was made between Kartagener's syndrome and defects in the motor protein dynein. This defect results in immotility of cilia, such as those in the respiratory system, and of sperm flagella, which are required for motility and fertilisation. 

However, the connection between cilia and asymmetric development was not fully established for another twenty years. In 1994, Sulik et al. \shortcite{Sulik94} reported on the development of a `node' structure on the ventral side of mouse embryos at 7--9 days post fertilisation (figure 1c, marked `VN'), on which primary cilia were expressed (figure 1d, e). Primary cilia are microscopic hair-like organelles found on virtually all cells in the body; they typically occur one-per-cell, their internal structure differing from that of mucus-propelling lung cilia, and sperm flagella, in that the central microtubule pair are absent (detailed in \opencite{Hirokawa09}).

In contrast to mucus-propelling cilia, primary cilia were generally thought to be non-motile. Sulik et al.
hypothesised that motility of the primary cilia in the node may be required for left-right symmetry-breaking, this being the reason that situs inversus occurs in patients with defective dynein. Nonaka et al. \shortcite{Nonaka98} subsequently revealed that these cilia were indeed motile, and in normal embryos perform a `whirling' motion that creates a right-to-left flow (figure 1d, f). Gene knockout mouse embryos were produced which did not assemble nodal cilia; these mutants did not break left-right symmetry correctly and did not survive. 

Many significant advances have been reported in the biological literature since; we restrict our attention to four experimental and four theoretical studies relevant to the fluid mechanics. Nonaka et al. \shortcite{Nonaka02} showed that artificial flow could reverse the situs of normal embryos, and direct the situs of mutants with immotile cilia. 
A number of questions were posed: firstly, how do whirling cilia create such a directional flow, and secondly, how is the flow converted to asymmetric development?

\begin{figure}
\begin{center}
$
\begin{array}{ll}
\mbox{\textsf{\textbf{a}}} & \mbox{\textsf{\textbf{b}}} \\
 \includegraphics[scale=0.68]{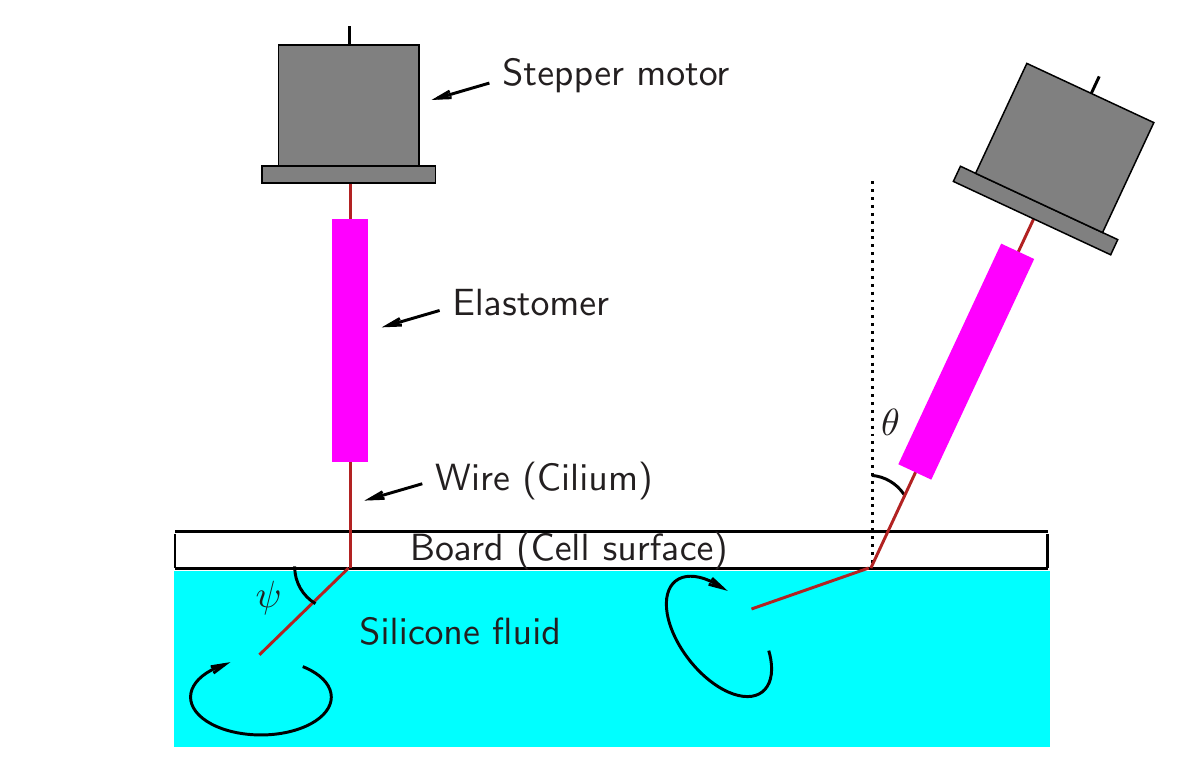} 
 & 
 \includegraphics[scale=0.36]{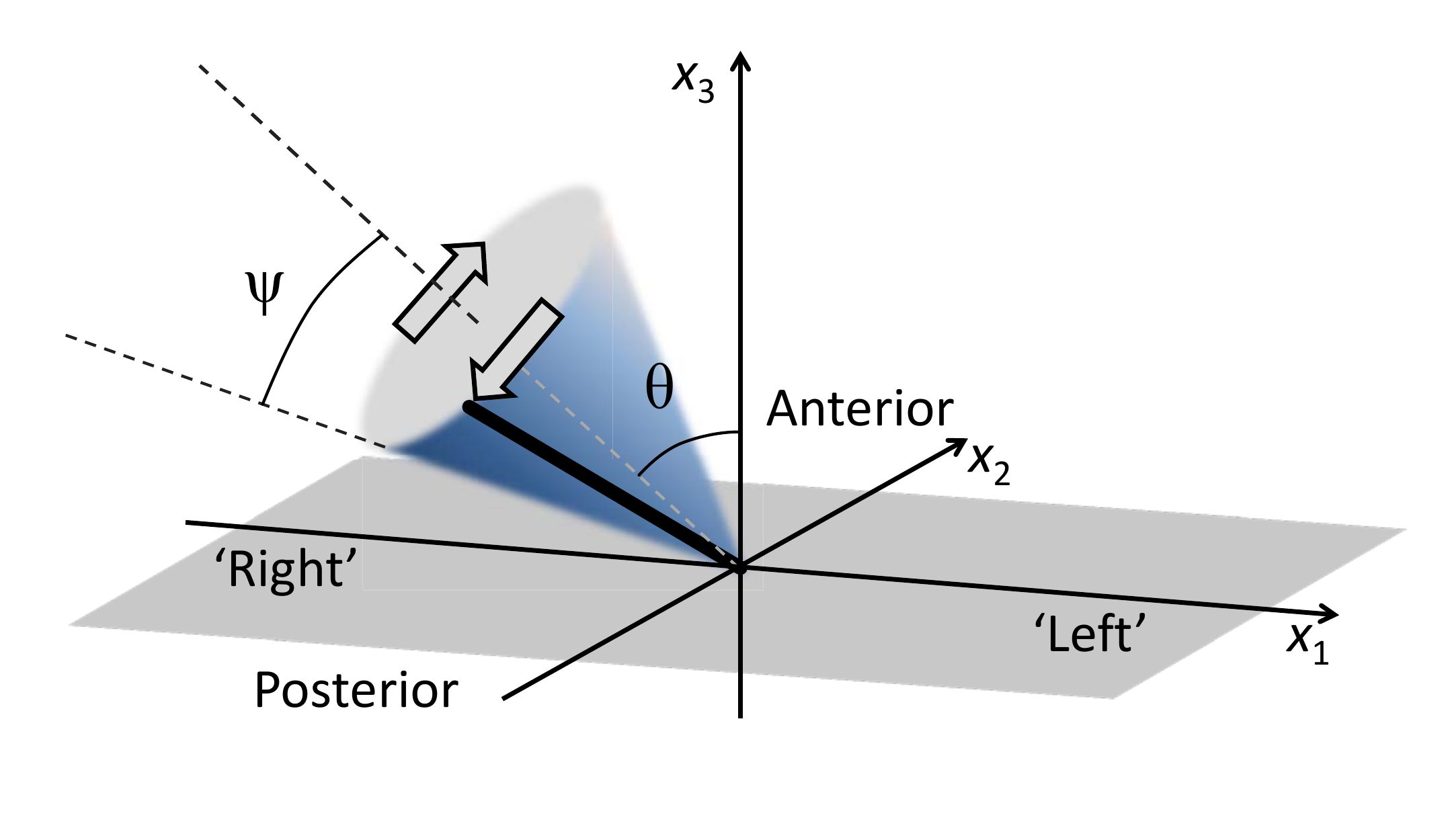}
\end{array}
$
\end{center}
\caption{(a) Schematic figure showing the experimental wire cilia model, redrawn from Nonaka et al. \shortcite{Nonaka05}. (b) Schematic showing the coordinate system used for the mathematical model of the conical rotation, its relation to the biological axes, and the conical rotation model for an individual cilium. The angle $\theta$ is the posterior tilt, $\psi$ is the semi-cone angle.}
\label{fig:nonakaWire}\label{fig:coneAngles}
\end{figure}

An answer to the first question was proposed by Cartwright et al. \shortcite{Cartwright04}. A fluid dynamic model using `rotlet' point-torque singular solutions of the Stokes flow equations was formulated, from which it was deduced that a posterior `tilt' in the rotational axis and the combined effect of multiple cilia arranged in an array could produce a right-to-left flow. Although this model did not take into account all of the essential physics of the cilium/surface interaction \shortcite{Smith08}, the `posterior tilt' was nevertheless a fundamental theoretical prediction, subsequently confirmed experimentally \shortcite{Okada05} (for more discussion of this work and the underlying modelling philosophy, see \shortcite{Cartwright08bdr}). Nonaka et al. \shortcite{Nonaka05} devised a different approach to investigate the fluid mechanics of nodal cilia: a laboratory analogue model using tilted rotating wires, with fluid viscosity sufficiently high and rotational speed sufficiently low to produce a flow with appropriately small Reynolds number (figure~\ref{fig:nonakaWire}a). 
A rotating rod tracing out a conical shape, tilted towards the posterior, can be defined with two parameters, the tilt angle, denoted $\theta$, and `semi-cone' angle, denoted $\psi$ (figure~\ref{fig:coneAngles}b).
Nonaka et al. observed that tilted rotation in their wire cilia model did indeed result in directional transport of marker particles; moreover they investigated how the tilt and semi-cone angles affected the particle transport rate. Optimal transport occurred when the angle sum was close to $90^\circ$, corresponding to the cilium being close to the epithelium during the return stroke; moreover a semi-cone angle of $60^\circ$ resulted in more rapid transport than semi-cone angle of $45^\circ$. The question of optimality of the cilia motion is discussed further in section~\ref{section:volumeFlowRate}.

The second question, regarding the conversion of the nodal flow to asymmetric development, was subject to two different hypotheses, one involving morphogen protein transport, the other mechanical sensing by cilia. An important breakthrough was made in the same year by Tanaka et al. \shortcite{Tanaka05}, who found that lipid-enclosed parcels of morphogen proteins, that affect cell differentiation and development, are released from the nodal cells and pushed into the flow field by microvilli. These `nodal vesicular parcels' (NVPs, figure 1f) were hypothesised to act as vessels for the transport of morphogens selectively to the left side of the embryo. The transport and break-up of these parcels within an enclosed volume was subsequently modelled as a steady flow in an enclosed domain, and solved numerically by the finite element method by Cartwright et al. \shortcite{Cartwright07}.

Our first study \shortcite{Smith07} was designed to address two important physical effects: the role of cilium-cell surface interaction, and the effect of unsteadiness in the flow field. The methodology, reviewed in this paper, was based on a numerical
application of slender body theory to calculate time-varying solutions to the Stokes flow equations, modelling the viscous-dominated microscale domain.
While these equations have no explicit time-dependence, the whirling motion of the cilia does have an explicit time-dependence, producing an unsteady flow.
An alternative explanation for nodal flow generation has been given in the literature, involving velocity differences in the effective and recovery stroke. However, due to the lack of explicit time-dependence of the Stokes flow equations, a motion performed in opposite directions with different speeds will not produce any net transport \shortcite{Childress81}, a fact sometimes referred to as the `Scallop Theorem'. 
By contrast, the unsteady boundary conditions and resulting unsteady flow have important implications for the transport of particles within the node.

Our computational simulations \shortcite{Smith07} investigated the flow produced by a small array of tilted cilia projecting from a plane in a semi-infinite fluid. This allowed calculation of three-dimensional particle transport under the unsteady flow field, and showed that `chaotic' effects could occur, specifically that particles close to the cilia beat envelopes, that are initially close together, could diverge considerably. This prediction was subsequently confirmed experimentally in zebrafish embryos using a novel laser ablation technique (\shortcite{Supatto08}, see section~\ref{section:physics} for more detail).
The results showed that an 'ensemble effect' is not required to produce a directional flow; a single cilium can produce a directional flow through an image-induced stresslet far-field, as described in detail in section~\ref{images}. 
The approach of using singular solutions and slender body theory was developed further in \shortcite{Smith08}, wherein an analytical formula was developed for the volume flow rate, closely paralleling the experimental findings of Nonaka et al. \shortcite{Nonaka05}; this is described in section~\ref{section:volumeFlowRate}.
Unsteady simulations of flow due to relatively large arrays of cilia were also presented; however 
a limitation of this work is that it did not account for the effect of the upper Reichert's membrane (figure 1f) on the flow field.
This had previously been accounted for in the steady flow model of Cartwright et al. \shortcite{Cartwright07},
who reported two separate regions of return flow: an upper return flow some distance above the cilia tips (figure 1f), and an additional lower region close to the epithelium (not shown), driven by the return stroke of the cilia. It therefore remains to investigate the effect of the membrane in the context of a time-dependent model. We shall first review in detail the anatomy and geometry of the node, then describe the mathematical modelling, and finally present new results for time-dependent flow with the upper membrane taken into account. 

\section{Anatomy of the node and geometric model}
The most commonly-studied species is the mouse; the essential details apply to organising structures found in the majority of species studied. 
Again, for detailed review, we refer to Hirokawa et al. \shortcite{Hirokawa09}. Briefly, the node of the mouse is an approximately triangular
depression that appears on the ventral side with the triangle apex pointing towards the anterior (figure 1b, c), 50--100 $\mu$m in width and 
10--20 $\mu$m in depth, covered with Reichert's membrane, and filled with extraembryonic fluid. The ventral surface, forming a tissue termed 
`epithelium', has a few hundred `nodal pit cells', generally expressing a single cilium of length 3--5 $\mu$m and diameter 0.3 $\mu$m, giving 
a slenderness ratio $O(1/10)$. The characteristic rotation rate has been quoted as approximately 600 rpm \shortcite{Hirokawa09}, or 10 Hz. 
The `nodal flow' first reported by Nonaka et al. \shortcite{Nonaka98} was visualised as the leftward movement of tracer particles at 2--5 $\mu$m/s 
at around 5 $\mu$m above the cell surface; a slower `return flow' occurs closer to the overlying membrane (\shortcite{Cartwright04,Okada05}; figure 1f).

To model the flow mathematically, we represent the epithelium surface by the plane $x_3=0$, the cilia extending into the region $x_3>0$, with 
the positive $x_1$ direction corresponding to the `left' and the positive $x_2$ direction corresponding to the anterior. 
The cilia are therefore tilted towards the negative $x_2$ direction. The cilia motion is modelled as a rod moving through a conical 
envelope; the cone has semi-cone angle $\psi$ and tilt angle $\theta$. 
The cilium cannot penetrate the epithelium surface, hence $\psi+\theta<90^\circ$. This is summarised in figure~\ref{fig:coneAngles}b.
Experimental data reported by Okada et al. \shortcite{Okada05} of the variation in the angles $\theta$ and $\psi$, and additionally angle of the tilt axis from the anterior, denoted $\phi$, are given in figure~\ref{fig:okada}(a, b, d) for
mouse, rabbit and medakafish. Additionally, values of the apparent cilium length $\rho$ are given (figure~\ref{fig:okada}c), showing considerable variation even within the class mammalia.
The rotation rate, typically approximately $10$ Hz (figure~\ref{fig:okada}e), varies by approximately 20\% over a beat cycle in the mouse,
slowing slightly during the recovery stroke, during which surface-induced viscous drag will be greatest.

In the sections below we describe the mathematical theory, starting with the fundamental singularities of Stokes flow, and the image singularities associated with
enforcing the no-slip condition on the cell surface. We then describe the use of slender body theory, and review the derivation of an analytic formula estimating
volume flow rate due to a single cilium. We then discuss the physical aspects arising from the image systems and slender body theory in more detail, and how
their predictions have subsequently been confirmed by experimental observations. We then apply the regularized stokeslet method to model Reichert's membrane for
the first time, and report computational results showing how this modifies the flow fields and particle tracks. We finish by relating these simulation
results to the biology of left-right symmetry-breaking, and outline future work for fluid mechanics and morphogen diffusion studies.

\begin{figure}
\begin{center}
\end{center}
\caption{Experimental data of cilia distribution and kinematics in three species, mouse, rabbit and medakafish (copyright figure not included in arxiv version) (a) Schematic of the tilt angle $\theta$, semi-cone angle $\psi$, angle of cilium rotation axis from anterior $\phi$, cilia projected length $\rho$ and
rotation angle $\xi$---the latter three scalar variables are used only for this figure and are not explicitly referred to in the mathematical model.
Axis directions are marked A, P (anterior-posterior) and L, R (left-right). 
(b) Data for tilt angle $\theta$ versus angle of rotation axis from posterior, $\phi$---the approximate mean of $180\degree$ corresponds to a 
complete posterior tilt. (c) Frequency distribution of cilium projected length. (d) Frequency distribution of semi-cone angle, inferred 
from the shape traced out by the cilium tip during the beat cycle. (e) Data showing the rotation angle traced out over time by a 
population of cilia in the mouse node, with the mean rotation over one beat cycle shown.
}
\label{fig:okada}
\end{figure}

\section{Singularity methods for fluid mechanics modelling}
The mathematical techniques in this paper will be based on singular solutions of the zero Reynolds number fluid flow equations, and their regularized counterparts. Firstly we review these solutions and how they give insight into the nodal flow, before describing the associated slender body theory. 

\subsection{Fundamental singularities of Stokes flow}
Due to the very small length and velocity scales, the Reynolds number of the nodal flow is much smaller than one, and so the Stokes flow equations are a very accurate model for the fluid dynamics. In this section we review a number of `fundamental solutions', representing the flow due to concentrated force, torque, strain and a source dipole. The linearity of the Stokes flow equations allow these solutions to be summed or integrated to satisfy boundary conditions associated with the epithelium or Reichert's membrane, and the flow produced by the movement of the cilia. Hence the properties of these solutions may be used to understand the physical mechanisms generating the flow, and also as a basis for slender body theory and boundary integral methods, which allow efficient and accurate computational modelling.

The Stokes flow equations with a force per unit volume $\bm{F}$ centered at $\bm{y}$ are,
\begin{equation}
\nabla p = \mu\nabla^2\bm{u} + \bm{F}\chi(\bm{x} - \bm{y}) \mbox{,} \quad \quad \nabla\cdot\bm{u} = 0 \mbox{,}
\label{eq:stokes-1}
\end{equation}
where $p=p(\bm{x},t)$ is pressure, $\bm{u}=\bm{u}(\bm{x},t)$ is velocity and $\mu$ is dynamic viscosity.
The term $\chi(\bm{x}-\bm{y})$ may be the Dirac delta distribution, representing a singular point-force, or a smoothed `blob' function (section~\ref{section:integralEq}). With $\chi(\bm{x}-\bm{y})$ taken as the delta distribution $\delta(\bm{x}-\bm{y})$, the solution to these equations is given by the stokeslet tensor, which in an infinite fluid takes the form
\begin{equation}
S_{ij}(\bm{x},\bm{y}) = \frac{1}{8\pi\mu}\left(\frac{\delta_{ij}}{r} + \frac{r_ir_j}{r^3}\right) \mbox{,}
\label{eq:stokeslet-1}
\end{equation}
$r_i = x_i - y_i$ $(i = 1,2,3)$ and $r^2=r_1^2+r_2^2+r_3^3$; the velocity then being given by $u_i=\sum_{j=1}^3 S_{ij}F_j$. Henceforth the summation convention over repeated indices will be assumed. Taking $\chi$ to be a smoothed blob function leads to a regularized stokeslet solution, introduced by \shortcite{Cortez01}. The infinite domain stokeslet decays with $O(1/r)$ in the far-field.

As described in detail in Blake \& Chwang \shortcite{Blake74a}, the stokeslet singularity may be differentiated to give a stokes-doublet singularity, a rank 3 tensor. This can be decomposed into symmetric and antisymmetric terms, referred to as the `stresslet' and `rotlet' singularities respectively
\begin{equation}
S_{ijk}^{D}(\bm{x},\bm{y}) = \frac{1}{8\pi\mu}\left[\left\{-\frac{r_i}{r^3} + \frac{3r_ir_jr_k}{r^5}\right\} + \left\{\frac{r_k\delta_{ij} - r_j\delta_{ik}}{r^3}\right\}\right] \mbox{.}
\label{eq:stokes-doublet-1}
\end{equation}
The first term is a rank 3 tensor, symmetric in $j,k$, termed the `stresslet'; when multiplied by a rank 2 tensor giving the stresslet strength, it generates a velocity field due to a concentrated straining motion of the fluid. The second term, antisymmetric in $j,k$, termed the `rotlet', can  be shown to be a rank 2 tensor; when multipled by a vectorial strength, it generates the velocity field due to a concentrated torque applied to the fluid. The rotlet can be rewritten as, $L_{ij}(\bm{x}-\bm{y})=\varepsilon_{ijk}r_k/(8\pi \mu r^3)$, the velocity field being given by $u_i=L_{ij}\Omega_j$ for point-torque $\Omega_j$.
Both the stresslet and rotlet velocity fields decay with $O\left(1/r^2\right)$ in the far-field, more rapidly than the stokeslet.

Another relevant singular solution of the Stokes flow equations is the point source dipole, a special case of a stokes-quadrupole, which tensorially is given by,
\begin{equation}
D_{ij}(\bm{x},\bm{y}) = -\frac{1}{4\pi}\left(\frac{\delta_{ij}}{r^3} -\frac{3r_i r_j}{r^5}\right) \mbox{.}
\label{eq:sourceDipole-1}
\end{equation}
This can be derived by differentiating the stresslet, then multiplying by the idemfactor $\delta_{jk}$.
The source dipole decays more rapidly than the stresslet or rotlet, being $O\left(1/r^3\right)$. This singular solution is used in both the construction of no-slip image systems in section~\ref{images}, and the formulation of slender body theory (section~\ref{section:sbt}).

\subsection{From fundamental singularities to nodal flow}\label{images}
An array of infinite domain rotlets $L_{ij}$ was used by Cartwright et al. \shortcite{Cartwright04}, inspiring their prediction of the posterior tilt; it was shown that an array of rotlets produces a unidirectional flow in the region spanned by the singularities. This model does not however satisfy the no-slip condition on the epithelium; moreover it predicts that the far-field flow is rotational. In this section we review the work of Blake \shortcite{Blake71a} and Blake \& Chwang \shortcite{Blake74a} on the use of image singularities to satisfy the no-slip conditions, and the implications of this for the nodal flow.

To model cilia projecting from a cell surface, Blake \shortcite{Blake71a} derived the following solution for a point-force a height $h$ above the no-slip surface $x_3=0$,
\begin{equation}
B_{ij}(\bm{x},\bm{y}) = \frac{1}{8\pi\mu}\left[\left(\frac{\delta_{ij}}{r} + \frac{r_ir_j}{r^3}\right) - \left(\frac{\delta_{ij}}{R} + \frac{R_iR_j}{R^3}\right) + 2h
\Delta_{jk}
\frac{\partial}{\partial R_k}\left\{\frac{hR_i}{R^3} - \left(\frac{\delta_{i3}}{R} + \frac{R_iR_3}{R^3}\right)\right\}\right] \mbox{,}
\label{blakeimages}
\end{equation}
where the tensor $\Delta_{jk}$ takes value $+1$ for $j=k=1,2$, value $-1$ for $j=k=3$, and zero if $j\neq k$, originally written as $(\delta_{j\alpha}\delta_{k\alpha}-\delta_{j3}\delta_{k3})$ with $\alpha=1,2$ by Blake \shortcite{Blake71a}. The image location is given by $R_1=r_1$,
$R_2=r_2$ and $R_3=-h$.

\begin{figure}[h]
$
\begin{array}{ll}
\includegraphics[scale = 0.32]{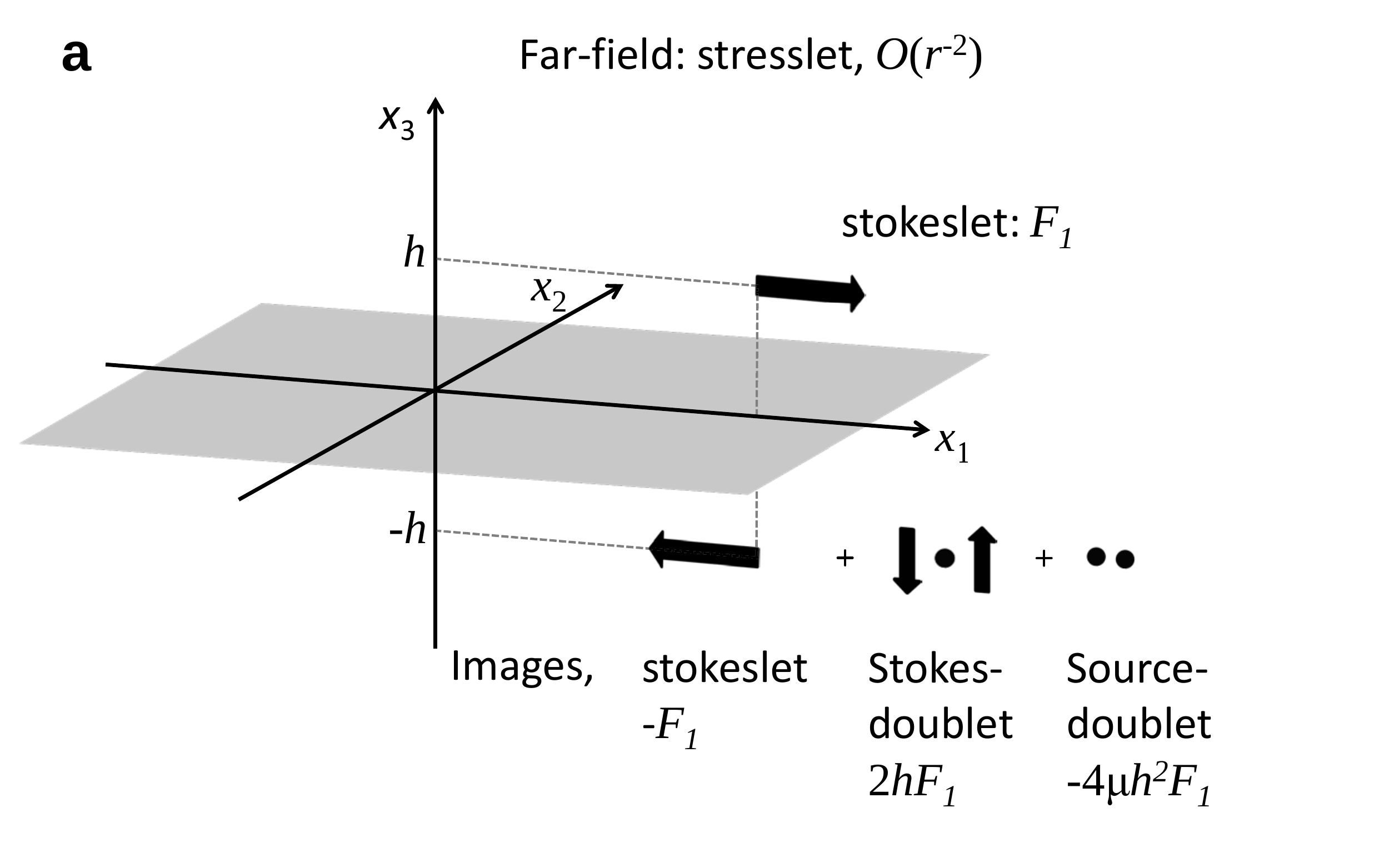} 
&
\includegraphics[scale = 0.32]{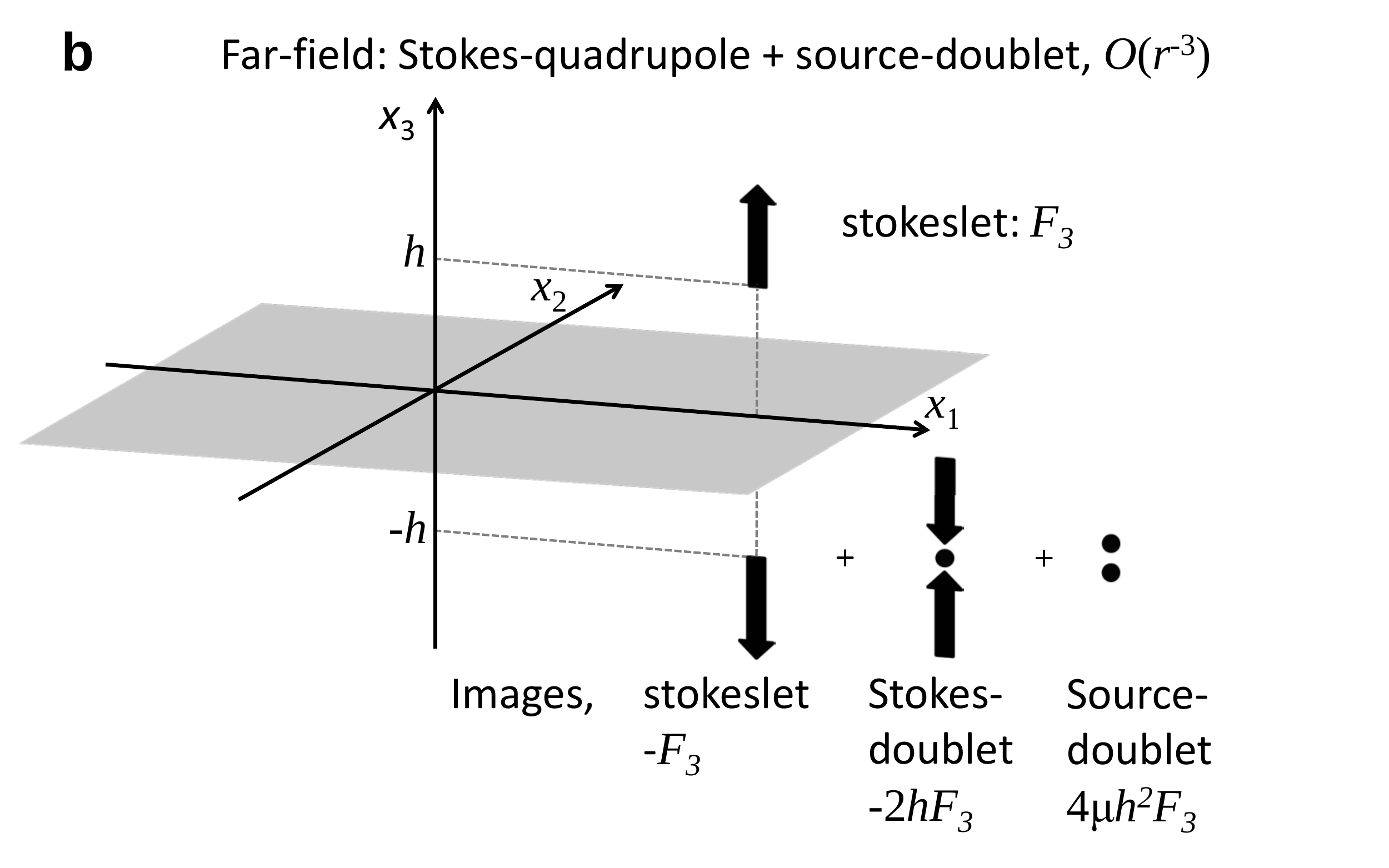} \\
\end{array}
$

$
\begin{array}{ll}
\includegraphics[scale = 0.32]{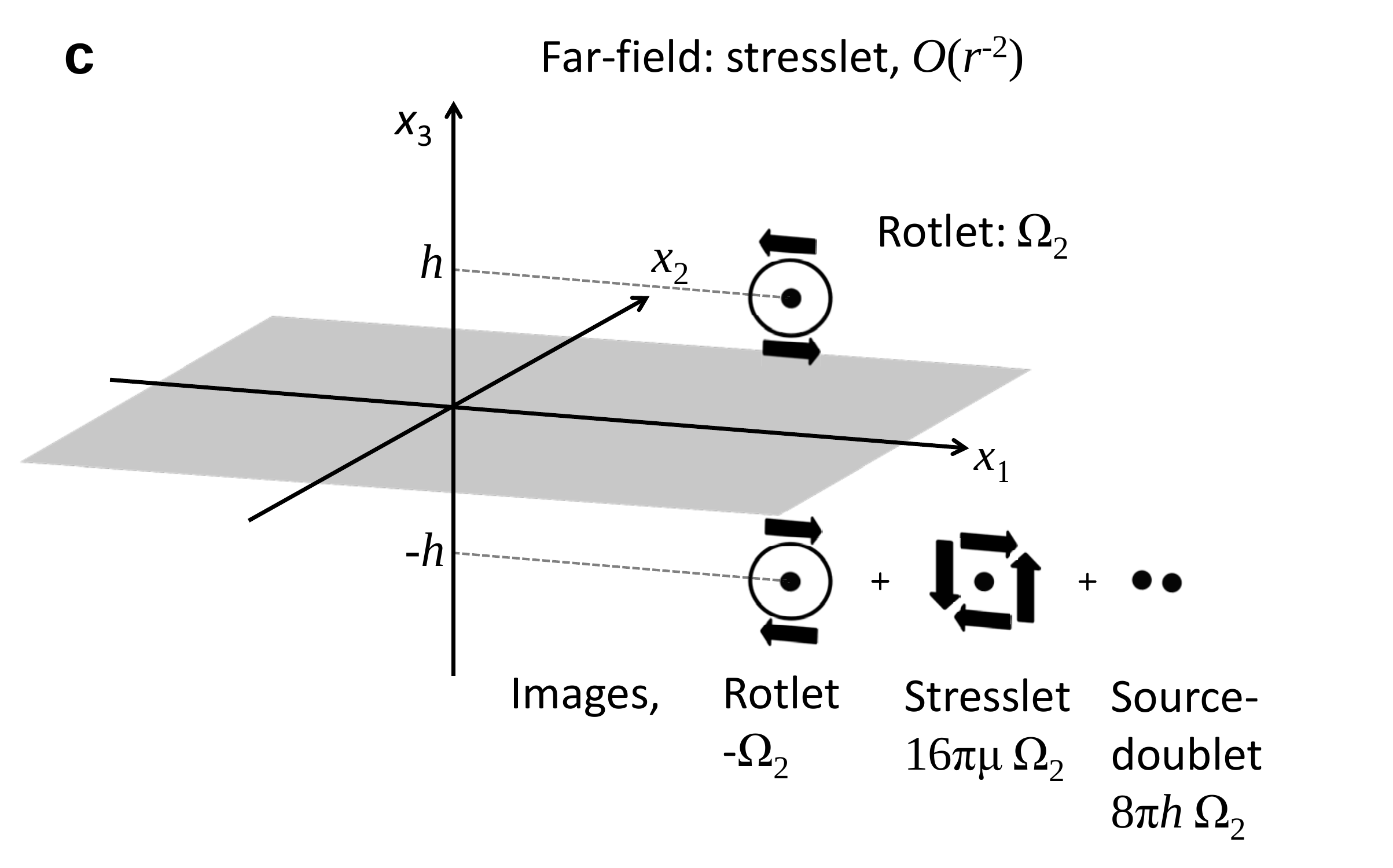} 
&
\quad
\includegraphics[scale = 0.32]{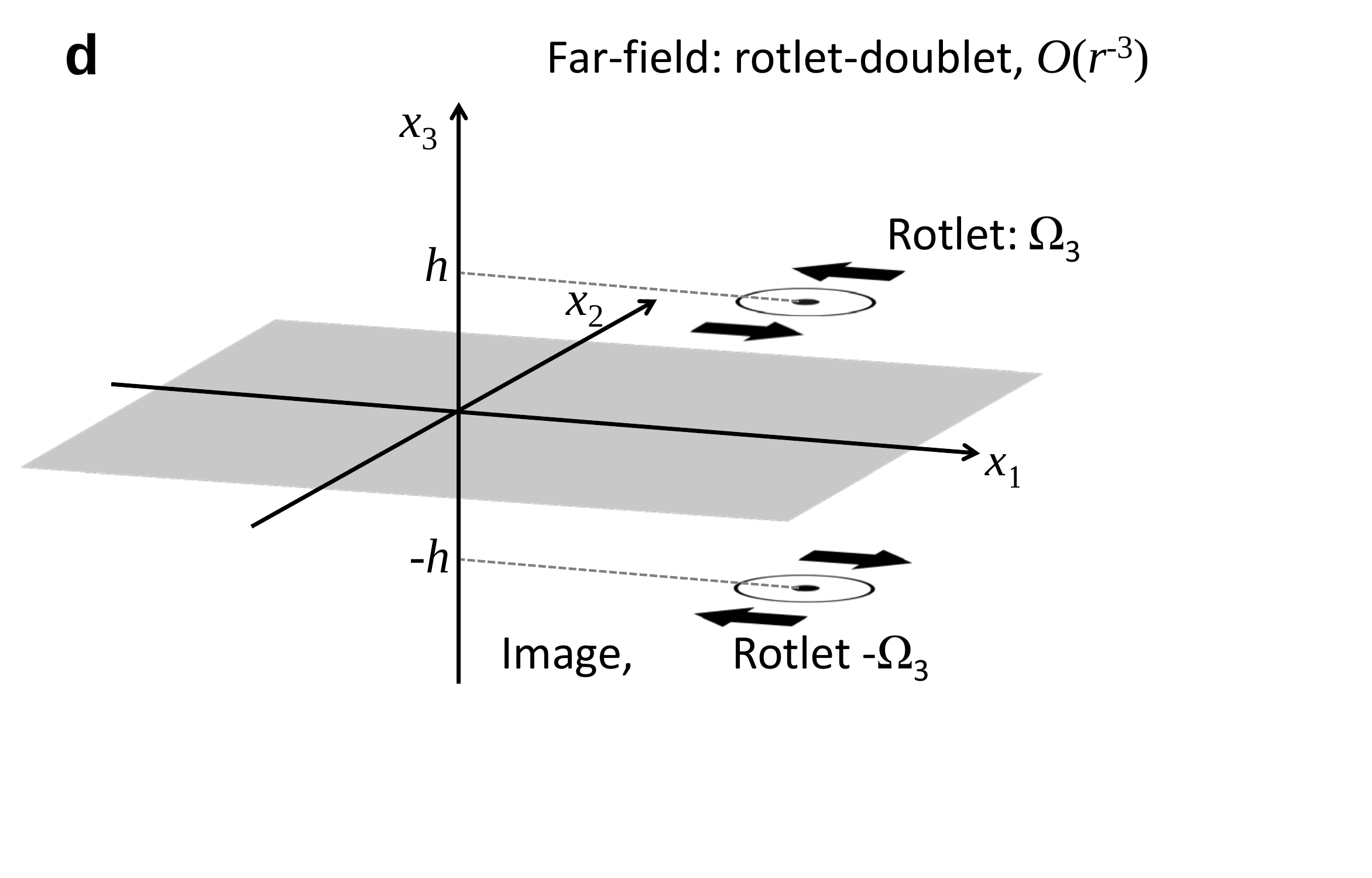} \\
\end{array}
$
\caption{Diagrams illustrating the image system for stokeslets $(a)$ $j = 1$ and $(b)$ $j = 3$, the strengths of the components being given in brackets. Note the significant difference in the far-fields; a stresslet for $j = 1\textrm{ or }2$ and a lower order and weaker stokes-quadrupole for $j = 3$. $(c, d)$ The image system and far-fields for a rotlet with components (c) tangential and $(d)$ normal to the wall.}\label{fig:imageSystems}
\end{figure}

The image system can be analysed as shown in figure~\ref{fig:imageSystems}; briefly it consists of an equal and opposite image stokeslet, a stresslet and a source doublet. The far-field of $B_{ij}$ is that of a stresslet, and implies that the direction of the streamlines will asymptote to being parallel to the vector $\bm{x}-\bm{y}$. The 
streamlines of the stresslet point out radially in the far-field, which was evident in the particle tracking simulations of Smith et al. \shortcite{Smith08}. Moreover, the stokeslet decay rate of $O(1/r)$ is converted to $O\left(1/r^2\right)$ by the image system. The stokeslet and image system have been used for an array of problems both within and outside of biological fluid mechanics; for the problem considered they form the basis of a slender body theory model of cilia protruding from a cell surface, as described in section~\ref{section:sbt}. 

The rotlet singularity also has an associated image system, first given by Blake \& Chwang \shortcite{Blake74a},
\begin{equation}
L_{ij}^{I}(\bm{x},\bm{y}) = \frac{1}{8\pi\mu}
\left[\frac{\varepsilon_{ijk}r_k}{r^3} - \frac{\varepsilon_{ijk}R_k}{R^3} + 2h\varepsilon_{kj3}\left(\frac{\delta_{ik}}{R^3} - \frac{3R_iR_k}{R^5}\right) + 6\varepsilon_{kj3}\frac{R_iR_kR_3}{R^5}\right].
\label{eq:rot-image}
\end{equation}
Of note is that the far-field is similar to that of the stokeslet singularity, being a stresslet; a rotational near-field motion is `converted' to a far-field straining motion by the influence of the surface. 

Stokeslets combined with their image system have generally been used for time-dependent models of cilia-driven nodal flow using slender body theory, as in \shortcite{Smith07}. This approach will be described in more detail in section~\ref{section:sbt}.
The rotlet with no image system has been used as a basis for a time-averaged model \shortcite{Cartwright04}; from the above physical arguments we observe that once the rotlet image system is taken into account, the two approaches are consistent in the far-field flow they produce. For further discussion of the consistency of these approaches, see Smith et al. \shortcite{Smith07}.

\subsection{Slender body theory}\label{section:sbt}
Slender body theory represents the flow driven by one or many cilia as centreline
distributions of stokeslet and higher order singularities. For review and computational implementation, see \shortcite{Smith07}, for asymptotic justification, see \shortcite{Johnson80} and for an analytic approach to the modelling of bodies with varying diameter axisymmetric geometries, see Blake et al. \shortcite{Blake10}. The essence of these methods is that the high aspect ratio of the cilium radius to length $a/L$ allows both near-field and far-field flows to be accurately approximated by a line distribution,
\begin{equation}
u_i(\bm{x},t) = \int_0^1 \left[B_{ij}(\bm{x};\bm{\xi}(\ssv,t))f_j(\ssv,t)  + D_{ij} (\bm{x};\bm{\xi}(\ssv,t))g_j(\ssv,t) \right] \; \mathrm{d}\ssv \mbox{,} \label{eq:sbt}
\end{equation}
where $f_j(\ssv,t)$ is force per unit length on the cilium, and $g_j(\ssv,t)$ is a source-dipole distribution strength. The task is then to determine strengths of these line distributions that will satisify as closely as possible the boundary condition $u_i(\bm{x})=\partial_t \xi_i(s,t)$, where $\bm{x}$ is any point on the cilium surface around the centreline point $\bm{\xi}(\ssu,t)$. The curved surface of the cilium can, for example, be represented conveniently as a curved cylinder, $\bm{X}(\ssu,t;\alpha):=\bm{\xi}(\ssu,t)+a(\bm{n}(\ssu,t)\cos\alpha+\bm{b}(\ssu,t)\sin\alpha)$ where $0\leqslant\alpha<2\pi$ and $\bm{n}$, $\bm{b}$ are unit normal and binormal respectively. In this case, the appropriate dipole weighting, as originally used by Hancock \shortcite{Hancock53} is $g_j(\ssv,t)=-(a^2/4\mu)f_j^{\perp}(\ssv,t)$, where $f_j^{\perp}(\ssv,t)$ is the component of the force density projected onto the plane given by the normal and binormal. This model is reasonably accurate away from the cilium tip (see \cite[Chap 6]{Smith07,Childress81}); from it we can develop a simplified `resistive force theory' model, as first used by Gray \& Hancock \cite{Gray55}. A simple derivation follows:

Consider the velocity field on the cilium surface, and approximate the integral with its `local' part only, 
\begin{equation}
u_i(\bm{X}(\ssu,t;\alpha)) = \int_{|\ssv-\ssu|<q} \left[B_{ij}(\bm{X}(\ssu,t;\alpha);\bm{\xi}(\ssv))f_j(\ssv,t)  + \frac{a^2}{4\mu} D_{ij} (\bm{X}(\ssu,t;\alpha);\bm{\xi}(\ssv))f_j^{\perp}(\ssv,t)\right] \; \mathrm{d}\ssv \mbox{,} \label{eq:sbtloc}
\end{equation}
for an appropriately-chosen length $q$. Neglecting cilium curvature on the segment $(\ssu-q,\ssu+q)$, and assuming that the force density can be approximated by its midpoint value $f_j(\ssu,t)$, the integrals can be calculated analytically, as first given by Hancock \cite{Hancock53},
\begin{equation}
u_i(\bm{X}(\ssu,t;\alpha)) = (C_t)^{-1}f_j^{t}(\ssu,t) + (C_n)^{-1}f_j^{n}(\ssu,t) + (C_b)^{-1}f_j^{b}(\ssu,t)\mbox{,}
\end{equation}
where the `resistance coefficients' $C_t, C_n, C_b$ in the tangential, normal and binormal directions respectively can be calculated analytically. It is not possible to derive optimal values of these in general \cite{Lighthill96reinterp}, however for cilium motion, successfully-used values are
\begin{equation}
C_t=\frac{4\pi \mu} {-1+2\log(2q/a)} \mbox{,} \quad C_n=C_b=\frac{8\pi \mu} {1+2\log(2q/a)} \mbox{,} \quad
\end{equation}
with $q=(aL)^{1/2}$ \cite{Gueron92}.
For flagellar motions and `9+2' cilia motion, an important property of these coefficients is the anisotropy ratio $\gamma=C_n/C_t\approx 2$. For our model of a nodal cilium as a straight whirling rod, tangential motion is zero, and hence the values of $C_t$ and $\gamma$ are not relevant. This `resistive force theory' approach will be used in section~\ref{section:volumeFlowRate} to estimate the
volume flow rate produced by a single cilium. The formulae for $C_n$ and $C_b$ show that the force density on the cilium is $O(\omega L/\log(L/a))$ where $\omega$ is the angular frequency of rotation.

While this approach gives a good approximation to the cilium-fluid interaction, if accurate flow velocity fields are required, this method is limited in that is not possible to obtain a force distribution that satisfies $u_i(\bm{\xi}(\ssu,t))=\partial_t \xi_i(\ssu,t)$ uniformly towards the cilium ends. A more refined approach is
to use the quadratic dipole weighting $g_j(\ssv,t)=-a^2s(1-s) f_j(\ssv,t)$, based on the analytic solution of Chwang \& Wu \shortcite{Chwang75} for a straight rod, subsequently generalised to an asymptotic analysis by Johnson \shortcite{Johnson80} and numerical implementation by Smith et al. \shortcite{Smith07}. If the surface of the cilium is taken to be a curved slender ellipsoid, with
$\bm{X}(s,t;\alpha):=\bm{\xi}(s,t)+a [1-(s-1/2)^2/(a^2+1/4)]^{1/2} (\bm{n}(s,t)\cos\alpha+\bm{b}(s,t)\sin\alpha)$,
the singular solutions being distributed between the foci of the ellipsoid, $s=0, 1$,
then accuracy to $O\left(\left(a/L\right)^2\right)$ can be obtained uniformly along the cilium, including the ends. In section~\ref{section:results} we denote the combination of plane boundary stokeslets $B_{ij}$ and source dipoles $D_{ij}$ with weighting $g_j(\ssv,t)=-a^2s(1-s) f_j(\ssv,t)$ by the symbol $G_{ij}$.

\subsection{Volume flow rate}\label{section:volumeFlowRate}
To give an indication of the semi-cone and tilt angles that give optimal fluid transport, we consider the volume flow rate in the $x_1$ (`leftward') direction across the half-plane given by $x_3>0$ and $-\infty<x_2<\infty$, for fixed arbitrary $x_1$ produced by a line distribution of the 
form~(\ref{eq:sbt}). The source dipole $D_{ij}$ produces zero flow across this plane, while a unit stokeslet $B_{ij}$ located at $x_3=h$ pointing in the $x_j$ direction produces a volume flow rate of $h/(\pi\mu)$ for $j=1$ and zero otherwise \shortcite{Liron78}. 

The conical rotation can be parameterised as follows---with no posterior tilt,
\begin{eqnarray}
\xi_1(s,t) & = & s   \sin \psi \cos (\omega t) \mbox{,} \nonumber \\ 
\xi_2(s,t) & = & -s \sin \psi \sin (\omega t) )\mbox{,} \nonumber \\ 
\xi_3(s,t) & = & s \cos \psi \mbox{.}
\end{eqnarray}
Applying an anticlockwise rotation by angle $\theta$ about the $x_1$ axis, we have
\begin{eqnarray}
\xi_1(s,t) & = & s   \sin \psi \cos (\omega t) \mbox{,} \nonumber \\ 
\xi_2(s,t) & = & s (-\sin \psi \sin (\omega t) \cos\theta-\cos \psi \sin \theta )\mbox{,} \nonumber \\ 
\xi_3(s,t) & = & s (-\sin \psi \sin (\omega t) \sin\theta+\cos \psi \cos \theta) \mbox{.} \label{eq:conical}
\end{eqnarray}
The velocity $\bm{u}(s,t)$ of the cilium can be determined by taking the time derivative of $\bm{\xi}(s,t)$. 
Noting that $\bm{u}(s,t)$ is normal to the cilium centreline, we can define a unit normal vector $\bm{n}(t)$ 
such that $\bm{u}(s,t)=\bm{u}_n(s,t)=\omega s \bm{n}(t)$. Resistive force theory gives the approximation $f_1(s,t)= C_n \omega s n_1(t)$, with 
$n_1(t)=-\sin \psi \sin (\omega t)$. 

As discussed above, a point-force of strength $f_1$ pointing in the $x_1$ direction, situated at height $\xi_3$ above a no-slip surface, 
produces an instantaneous volume flow rate of $f_1 \xi_3/(\pi\mu)$; point-forces acting in the $x_2$ and $x_3$ directions produce zero flow. 
The volume flow rate induced by the force $C_n \omega s \bm{n}(t)$ is therefore equal to $C_n \omega s n_1(t) \xi_3(s,t)/(\pi\mu)$.
Integrating this function from $s=0$ to $L$, and averaging over a beat cycle, the mean volume flow rate can therefore be written,
\begin{equation}
\bar{Q} = \frac{C_n \omega^2}{2\pi^2\mu} \int_{0}^L\int_{0}^{2\pi/\omega} s  n_1(t) \xi_3(s,t) \; \mathrm{d}t \; \mathrm{d}s\mbox{.} 
\end{equation}
Using the above expressions for $\xi_3(s,t)$ and $n_1(t)$ and noting that the integrals of $\sin^2(\omega t)$ and $\sin(\omega t)\cos(\omega t)$ over one period are $2\pi/\omega$ and zero respectively, we find that the above integral evaluates to
\begin{equation}
\bar{Q} = \frac{C_n \omega L^3}{6\pi \mu} \sin^2\psi \sin \theta \mbox{,}
\end{equation}
as given in \shortcite{Smith08}.

The mean volume flow rate $\bar{Q}(\psi,\theta)$ is the product of two non-negative, non-decreasing functions, and therefore attains a maximum value when $\theta + \psi = 90\degree$, in agreement with the outcomes observed by Nonaka et al. \shortcite{Nonaka05}.
Further analysis shows that the maximum occurs at $\psi=\mathrm{arctan}\sqrt{2}=54.7\degree$ and $\theta=35.3 \degree$, ignoring the fact that the slender body model loses accuracy as the cilium closely approaches the surface. This occurs when $\theta+\psi=90\degree$, the surface drag in the real system becoming much larger than $C_n \bm{u}_n$.
A range of values of the tilt angle $\theta$ have been observed experimentally \shortcite{Okada05}, however the mean values in mouse, rabbit and medakafish are typically between $35$--$40\degree$ (figure~\ref{fig:okada}b).
There is more spread in the reported values of the semi-cone angle $\psi$ for these species, their means ranging from approximately $40$--$50\degree$ (figure~\ref{fig:okada}d).

As discussed in the introduction, the experimental model of Nonaka et al. \shortcite{Nonaka05} using mechanically-driven wire `cilia' (figure~\ref{fig:nonakaWire}a) produced the most effective transport when $\theta+\psi=90\degree$. Furthermore, comparing transport with semi-cone angles of $\psi=45\degree$ and $\psi=60\degree$, more effective transport occurred in the latter case, a result with which the functional form for $\bar{Q}$ is consistent \shortcite{Smith08}.

We now review in more detail how slender body theory and fundamental solutions explain the physical mechanisms underlying flow generation.

\subsection{Physics underlying the nodal cilium beat cycle}\label{section:physics}
The physical effects governing the fluid motion generated by the whirling nodal cilia can be interpreted through equation~\eqref{blakeimages} and 
figure \ref{fig:imageSystems}(a, b), which give the velocity fields due to a force (stokeslet) near a rigid boundary.
In figure \ref{fig:fields}(a, b) the `influence zones' are schematically illustrated for the nodal cilium in its effective stroke (upright orientation) and recovery stroke (near epithelial surface) based on the stokeslet model. The diagrams in figures \ref{fig:fields}(a, b) are adaptations of earlier diagrams for the motion of planar-beating `9+2' cilia presented in Blake \shortcite{Blake72,Blake85}, 
Blake and Sleigh \shortcite{Blake74b} and for mucous transport by Sleigh et al. \shortcite{Sleigh88}.

Figure \ref{fig:fields}a shows the key physical features of the effective stroke highlighting the three zones of influence:
\begin{figure}[h]
\begin{center}
$
  \begin{array}{ll}
    \mbox{\lba} &  \mbox{\quad\quad\; \lbb} \\  
    \includegraphics[scale = 0.3]{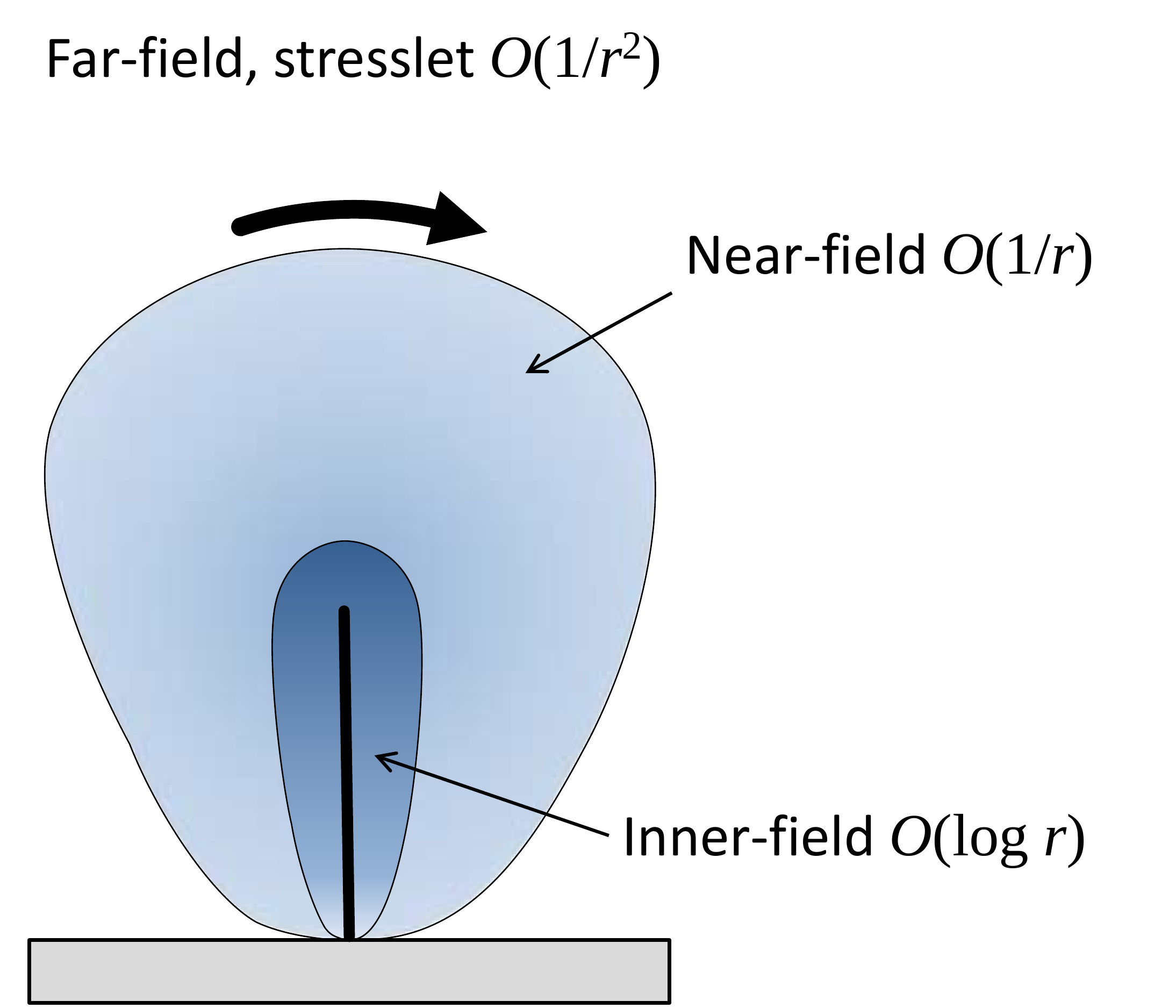}  
     &  \quad\quad \raisebox{20mm}{	$
       \begin{array}{l}   
         \includegraphics[scale = 0.3]{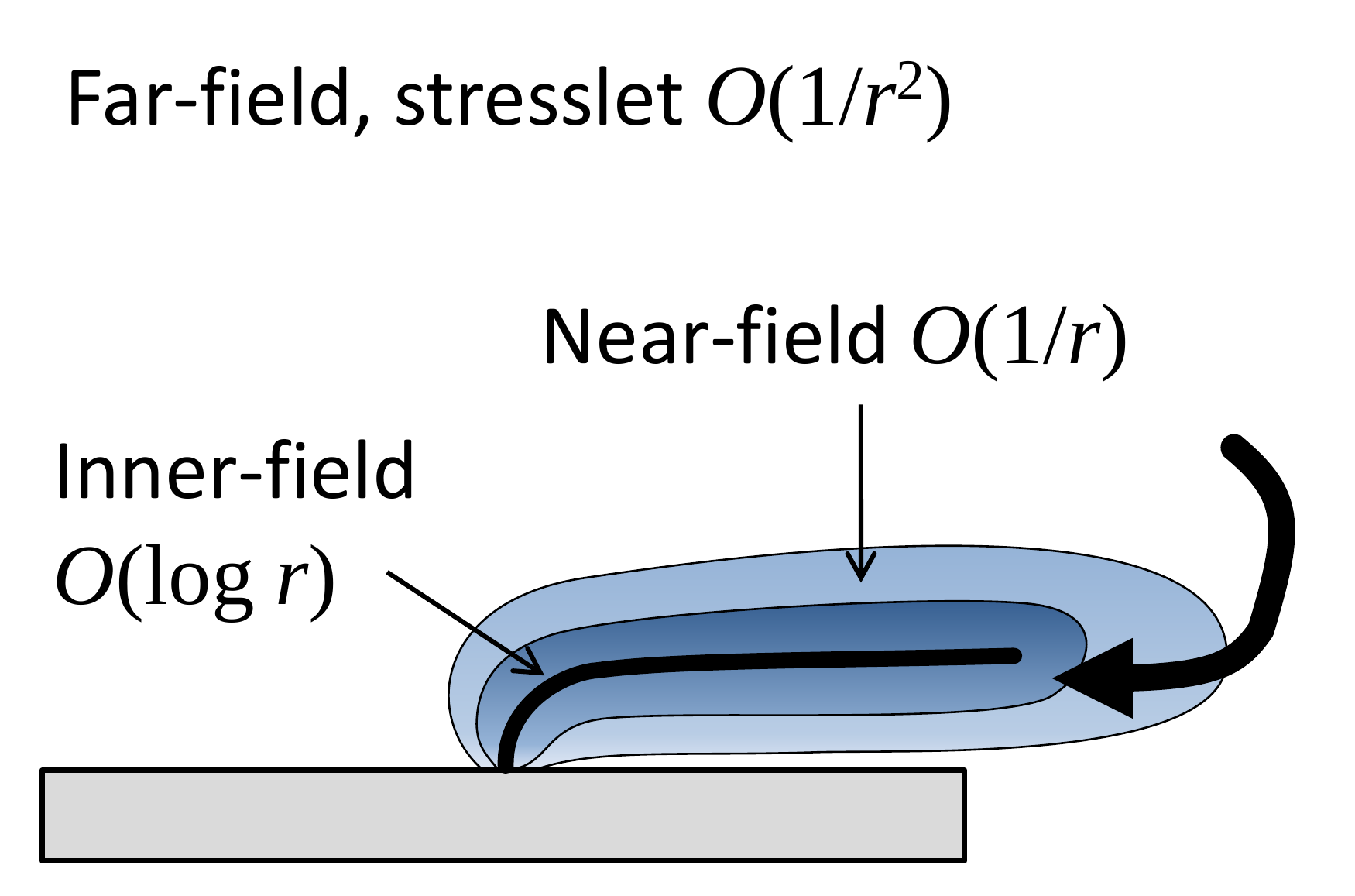}\vspace{6mm}\\
         \mbox{\lbc}\\
         \quad
         \includegraphics[scale=0.3]{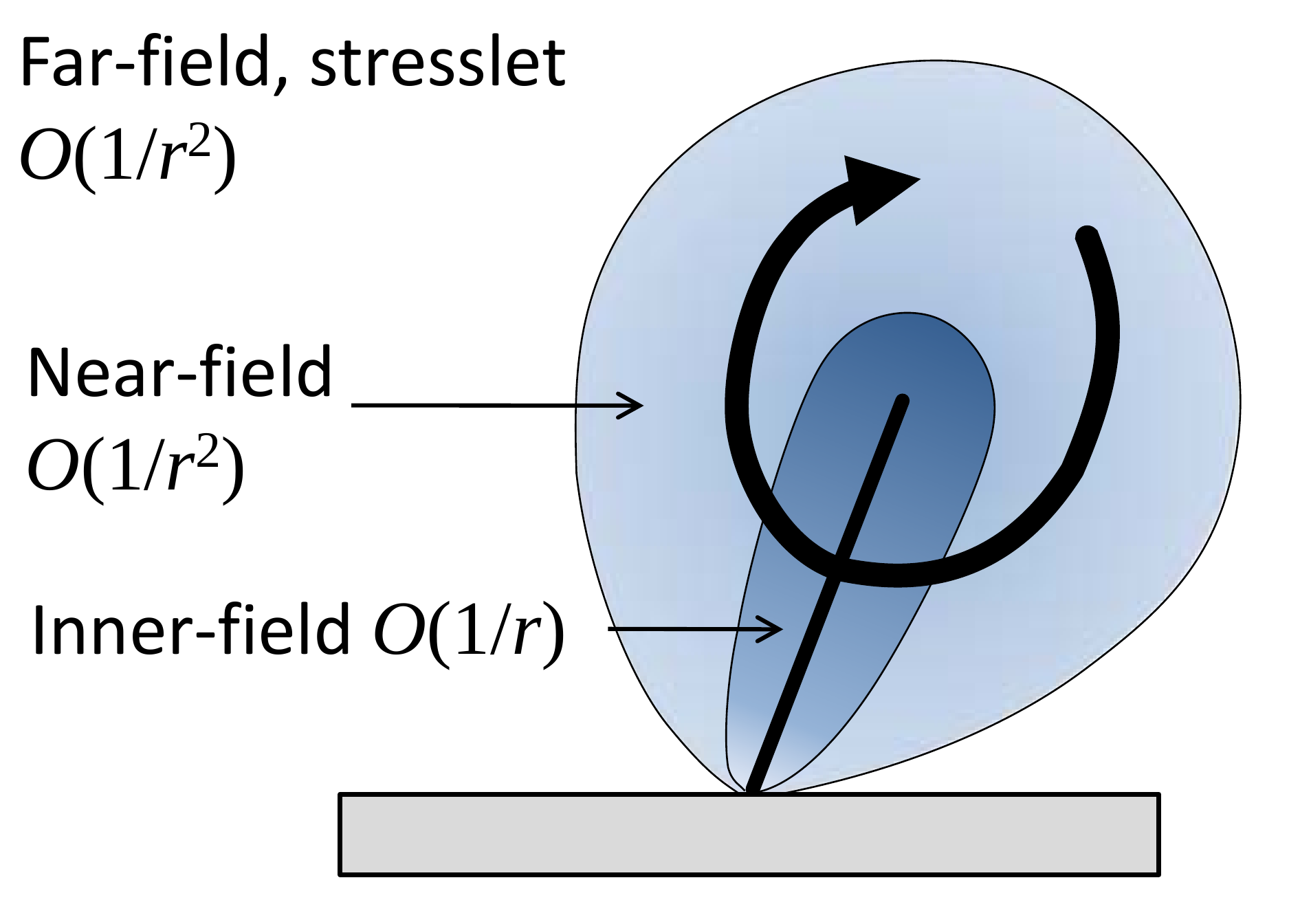}
       \end{array}
       $
       }
  \end{array}         
$
\end{center}
	\caption{The zones of influence of the cilium during the phases of its motion.
The three fields, `inner', 	`near' and `outer' decay differently with distance $r$ from the cilium. The inner-field corresponds to $r\ll L$, the near-field $r\sim L$ and far-field by $r \gg L$. (a, b) from the perspective of the instantaneous cilium motion, the relatively upright effective stroke (a) entrains a larger region of fluid than the relatively less upright recovery stroke (b). The inner-field is essentially given by a line distribution of stokelets and source-dipoles $({O}(\log r))$, the near-field by the stokeslet itself $({O}(1/r))$ and the far-field by the stokes-doublet image system $\left({O}\left(1/r^2\right)\right)$ shown in figure~\ref{fig:imageSystems}. (c) From the perspective of a time-averaged rotation, the inner-field is given by a line integral of rotlets $({O}(1/r))$, the near-field by the rotlet itself $\left({O}\left(1/r^2\right)\right)$ and the far-field by the stresslet image system $\left({O}\left(1/r^2\right)\right)$ shown in figure~\ref{fig:imageSystems}. The far-field velocity in all three cases is an $O\left(1/r^2\right)$ stresslet field.}
\label{fig:fields}
\end{figure}

\begin{figure}[h]
	\centering
	\caption{A depiction of the generation of the nodal flow (copyright figure not included in arxiv version). Cilia rotate anticlockwise, and due to their position on the curved cell surface, their rotational axis is tilted towards the posterior. 
This results in there being an effective leftward `propulsive' motion where the cilium projects above the cell surfaces, and a rightward `return stroke'  where the cilium is close to the cell surfaces. The viscous drag associated with the latter reduces the effect of the cilium during its return stroke, creating a flow from right to left.}
	\label{fig:generation_nodal_flow}
\end{figure}

\begin{itemize}
\item the inner-field---in the region for which $r\ll 1$, the scaled cilium length, the velocity field is ${O}(\log r)$, as shown by integrating stokeslet and
source-dipole distributions \shortcite{Lighthill76}.
\item the near-field---in the region for which $r={O}(1)$, the velocity field is ${O}(1/r)$, essentially dominated by the stokeslet field, although as $r$ becomes comparable with the distance from the image system $2h$, this will begin to have a significant effect.
\item the far-field---in the region for which $r\gg 1$, the stokeslet and image system are approximated by a symmetric stokes-dipole, and outcome of
the equal and opposite force of the wall and the couple term in the image system, giving a `stresslet' far-field, which decays with ${O}(1/r^2)$.
\end{itemize}

Figure \ref{fig:fields}b shows the features of the recovery stroke, which is close to the epithelium. The volume of fluid in the near-field is very greatly reduced
because the cilium is much closer to the epithelium, and hence the distance from the image system is comparable to $r$ even for relatively small values of $r$.

In the inner and near-fields, particles follow the rotational cilium motion; the near-field however shrinks during the rotational beat, so a significant volume of fluid
is subject to a relatively strong ${O}(1/r)$ near-field flow during the effective stroke, but a much weaker ${O}(1/r^2)$ far-field
flow during the return stroke. This leads to a `loopy drift' of particles towards the left. Further out, particles are in the stresslet far-fields of both the effective
and recovery strokes, however the former has a larger magnitude due to the stresslet strength being proportional to height above the surface. This leads to a time-averaged
stresslet far-field with radial streamlines, as evident in simulation results \shortcite{Smith07,Smith08}.
The predictions of a lower vortical flow and an upper directional flow were
subsequently confirmed by observations of the flow in Kupffer's vesicle of zebrafish embryos (\shortcite{Supatto08}, figure~\ref{fig:supatto}).

An alternate perspective is to represent the time-averaged rotational motion by rotlets with axis of rotation tilted towards the posterior \shortcite{Cartwright04}. Replacing the conical rotation by a line distribution
of rotlet singularities (figure~\ref{fig:fields}c), the inner-field has strength ${O}(1/r)$, as determined by integrating a rotlet distribution,
the near-field has strength ${O}(1/r^2)$, asymptotic to the strength of a rotlet, and the far-field, where the image singularities become equally important
has strength ${O}(1/r^3)$. Indeed, only the component of the rotlet parallel to the boundary will have a significant far-field effect, since a rotlet
perpendicular to the boundary decays with ${O}(1/r^3)$, as shown in figure~\ref{fig:imageSystems}(c, d). In the far-field, the rotlet has stresslet character, reconciling the two approaches for modelling the rotational cilium movement.

Figure~\ref{fig:generation_nodal_flow} from Hirokawa et al. \shortcite{Hirokawa09} summarises these findings: the posterior tilt of the nodal cilium results in a reduced flow rate during the recovery stroke compared with the effective stroke, which generates an overall flow. 

\begin{figure}
\begin{center}
\end{center}
\caption{A laser ablation particle tracking study of Kupffer's vesicle in zebrafish embryos proved the existence of the lower vortical flow and upper directional flow, in addition to chaotic particle transport (not shown) (copyright figure not included in arxiv version).}\label{fig:supatto}
\end{figure}

Our analysis does not however take into account the effect of the upper membrane (figure 1f), which will produce a `back pressure' gradient enforcing mass conservation, as first considered theoretically by Cartwright et al. \shortcite{Cartwright04}. 
In the next section we review the regularized stokeslet boundary integral method, and report new computational results applying this technique to the enclosed nodal cavity.

\section{The regularized stokeslet method}
\subsection{Integral equation}\label{section:integralEq}
For convenience in carrying out boundary integral calculations, and to give a regular flow field throughout the domain, including the regions containing singularity distributions, Cortez \citeyear{Cortez01} introduced the `regularized stokeslet'. This is defined as the exact solution to the Stokes flow equations with smoothed point-forces,
\begin{equation}
0  =  -\bm{\nabla}p+\mu\nabla^2\bm{u}+\bm{f}\psi_\epsilon(\bm{x}-\bm{\xi}) \mbox{,} \quad\quad  \bm{\nabla}\cdot\bm{u}   =  0        \mbox{.}
\end{equation}

\noindent The symbol $\psi_\epsilon(\bm{x}-\bm{y})$ denotes a cutoff-function or `blob' with regularisation parameter $\epsilon$, satisfying $\int_{\mathbb{R}^3} \psi_\epsilon(\bm{x})\mathrm{d}V_{\bm{x}}=1$. Cortez et al. \shortcite{Cortez05} showed that with the choice $\psi_\epsilon(\bm{x}-\bm{\xi}):=15\epsilon^4/(8\pi\mu r_\epsilon^7)$, the regularized Stokeslet velocity tensor is given by
\begin{equation}
S_{ij}^\epsilon(\bm{x},\bm{\xi})=\frac{\delta_{ij}(r^2+2\epsilon^2)+r_i r_j}{r_\epsilon^3} \mbox{.}
\label{regSto}
\end{equation}

\noindent Here and in the rest of the paper, we use the compact notation $r_\epsilon=\sqrt{r^2+\epsilon^2}$. 
We follow Cortez and co-authors in using the solution $S_{ij}^\epsilon$, and its counterpart near a no-slip boundary $B_{ij}^\epsilon$ given in equation~\eqref{ainleyimage}, as the basis for our computational study.

Ainley et al. \shortcite{Ainley08} derived the equivalent regularized image system, which we denote $B_{ij}^\epsilon$ for flow near a no-slip boundary, which we rewrite in index notation:
\begin{eqnarray}
B_{ij}^\epsilon(\bm{x},\bm{\xi}) & = & \frac{1}{8\pi\mu}\left(\frac{\delta_{ij}(r^2+2\epsilon^2)+r_i r_j}{r_\epsilon^3}   		                  
       - \frac{\delta_{ij}(R^2+2\epsilon^2)+R_i R_j}{R_\epsilon^3}       \right.        \nonumber \\
& &            + 2 h  \Delta_{jk} \left[ \pd{}{R_k} \left( \frac{h R_i}{R_\epsilon^3} -   \frac{\delta_{i3}(R^2+2\epsilon^2)+R_i R_3}{R_\epsilon^3} \right) -4\pi h \delta_{ik} \phi_\epsilon(R) \right] 
\nonumber \\
& &\left. -\frac{6h\epsilon^2}{R_\epsilon^5}(\delta_{i3}R_j-\delta_{ij}R_3) \right)
\mbox{.}
\label{ainleyimage}
\end{eqnarray}
The tensor $\Delta_{jk}$ is as defined for equation~(\ref{blakeimages}); $R_\epsilon^2:=R^2+\epsilon^2$ and the term $\phi_\epsilon(R):=3\epsilon^2/(4\pi R_\epsilon^5)$ is a more slowly-decaying blob than $\psi_\epsilon$, generating the regularised source dipole image.

Cortez et al. \shortcite{Cortez05} derived the equivalent Lorentz reciprocal relation, and hence a boundary integral equation. This is the basis for the `regularized stokeslet method' (RSM). Again using linearity of the Stokes flow equations and adding together the boundary integral for a surface $S$ of regularized stokeslets and an array of slender body integrals for the cilia, we have the following equation for the fluid velocity at location $\bm{x}$,
\begin{eqnarray}
u_i(\bm{x},t)&=&\int \!\!\! \! \int_S B_{ij}^\epsilon(\bm{x},\bm{\xi}) \Phi_j(\bm{\xi},t) \; \mathrm{d}S_{\bm{\xi}} +\sum_{m=1}^{M} \int_0^L G_{ij}(\bm{x},\bm{\xi}^{(m)}(s,t)) f_j^{(m)}(s,t) \; \mathrm{d}s
 \mbox{.}
 \label{eq:rsIntVel}
\end{eqnarray}
The unknowns $\Phi_j(\bm{\xi},t)$ and $f_j^{(m)}(s,t)$ denote the $j$-components of the stress on the membrane $S$ at $\bm{\xi}\in S$ and force per unit length at arclength $s$ on the $m$th cilium respectively, these being unknown a priori. As defined in section~\ref{section:sbt}, the kernel $G_{ij}$ is a combination of stokeslets and quadratically-weighted source dipoles chosen to give optimal accuracy on the cilia surfaces.

\subsection{Numerical implementation}
As described previously \cite{Smith09b}, we implement a constant-element method where the numerical quadrature of the kernel and discretisation of the unknowns are `decoupled'. The membrane is decomposed into elements $S[1],\ldots,S[N_S]$, on which the stress is approximated by $\Phi_j[1],\ldots,\Phi_j[N_S]$; the $m$th cilium is decomposed into elements $I^{(m)}[1],\ldots,I^{(m)}[N_C]$ on which the force per unit length is approximated by $f_j[1],\ldots,f_j[N_C]$. The discrete system at time $t$ is then given by
\begin{eqnarray}
u_i(\bm{x},t)
&=&
\sum_{q=1}^{N_S} \Phi_j[\nu]\int\!\!\!\!\int_{S[q]} B_{ij}^\epsilon (\bm{x},\bm{\xi}) \; \mathrm{d}S_{\bm{\xi}}
+
\sum_{m=1}^{M}\sum_{q=1}^{N_C} f_j[q]\int_{I[q]} G_{ij} (\bm{x},\bm{\xi}^{(m)}(s,t)) \; \mathrm{d}s
\mbox{,}
\label{discretesys}
\end{eqnarray}
where $\bm{\xi}_q$ is the approximate centroid of the surface element $S[q]$ and the arclength $s_q=(q-1/2)/N_C$.
This is then solved by collocation, applying equation~\eqref{discretesys} at $\bm{x}=\bm{x}^l$, for $l=1,\ldots,(M\times N_C+N_S)$ locations at the midpoint of each
cilium and the approximate centroid of each surface element, for more details, see \shortcite{Smith09b}.

The cilia are discretised with intervals $I[q]=((q-1)/N_S,q/N_S)$; 
$N_S=12$ constant-force density elements were used per cilium, and a membrane surface mesh with $6\times 10 \times 10=600$ constant-stress elements, with geometry based on the projected cube mesh used by Cortez et al. \shortcite{Cortez05} rearranged to a hemisphere and deformed using equation~\eqref{mesh_deform}.  The regularisation parameter for the surface mesh was taken as $\epsilon=0.01A$, where $A$ is the notional hemisphere radius, based on numerical tests reported previously \shortcite{Cortez05,Smith09b}.
Surface integrals were performed using $12\times 12$ or $4\times 4$ Gauss-Legendre quadrature, the more refined rule being used if the evaluation point was within 
a distance $1.5$ of the element centre. 
Cilia stokeslet integrals were taken using standard analytic formulae (see for example \shortcite{Smith09b}); weighted source dipole integrals were taken using Gauss-Legendre quadrature. For the latter, quadrature with $12$ points was used if the distance from the element
centre was less than $0.2$, quadrature with $4$ points was used if the distance was between $0.2$ and $1.0$, 
and the integrals were neglected at greater distances, due to their rapid $O\left((a/r)^3\right)$ decay.

For an array of $25$ cilia beating in synchrony, as depicted in figure~\ref{fig:geometry}, the discrete approximations $f_j^{(m)}[q]$ and $\Phi_j[\nu]$for $j=1,2,3$, cilium number $m=1,\ldots,25$, cilium interval $q=1,\ldots,12$ and surface element $\nu=1,\ldots,600$ results in $3\times (600+25\times 12)=2700$ scalar degrees of freedom at each timestep. Due to the absence of any explicit time-dependence in the Stokes flow equations, there was no coupling between timesteps; the calculations at each timestep were independent. The linear system was solved by $LU$-decomposition, as described in \shortcite[Chap 2.3]{Press92}. Moreover, it is only necessary to compute the force density and stress distributions
for a single beat cycle due to periodicity. Computations were performed using the BlueBEAR and Babbage1 clusters at the University of Birmingham. 

Once the discrete approximations to $f_j^{(m)}[q]$ and $\Phi_j[\nu]$ have been calculated, the velocity field at any point in the fluid can then be calculated using equation~\eqref{discretesys}. To give a regular velocity field throughout the flow domain, the cilia slender body integrals being taken using regularized rather than singular kernels, the regularisation parameter $\epsilon$ being taken to be the minimum value of the radius $a(s)$ on each segment. The accuracy of this approximation was verified by computing the cilium surface velocity, as described in section~\ref{section:sbt}. Particle tracking was performed using a predictor-corrector 2nd order algorithm, as described previously \shortcite{Smith07}.

\section{Results}\label{section:results}
As in figure~\ref{fig:coneAngles}, all results are plotted in the conventional manner, with the `left' of the embryo on the right hand side of the figure, in the direction of the positive $x_1$ axis, and flow to the `left' of the embryo will be referred to as positive. The $x_2$ axis is then posterior to anterior and the $x_3$ axis is dorsal to ventral.

\subsection{Nondimensionalisation}
Results are presented using the following scalings: length is scaled with respect to cilium length, 
which in the mouse embryo is typically $3$--$5\;\mu$m. 
Time is scaled with respect to beat period $2\pi/\omega$, typically $0.1$ s for a beat frequency of $10$ Hz. 
Velocity is scaled accordingly, giving a characteristic velocity scale of $30$--$50\;\mu $m/s. 
The dimensionless cilium tip speed is therefore $2\pi\sin\psi$, typically around $130$--$220\; \mu$m/s. 

These scalings entail that the correct scaling 
for the force per unit length $f_j$ is $\mu\sigma L$, and for stress $\Phi_j$ is $\mu\sigma$, where $\mu$ is dynamic viscosity.
Viscometry data on the extraembryonic liquid do not appear to have been presented; experiments are typically carried out using saline
media with added serum, the viscosity being within an order of magnitude of that of water, so $\mu$ can be estimated as $0.001$ Pa s.

\subsection{Geometry}
\begin{figure}
$
\begin{array}{cc}
\mbox{\lba `Top' view, with mesh}           &  \mbox{\lbb `Side' view, mesh outline only}\\
\includegraphics[scale=1]{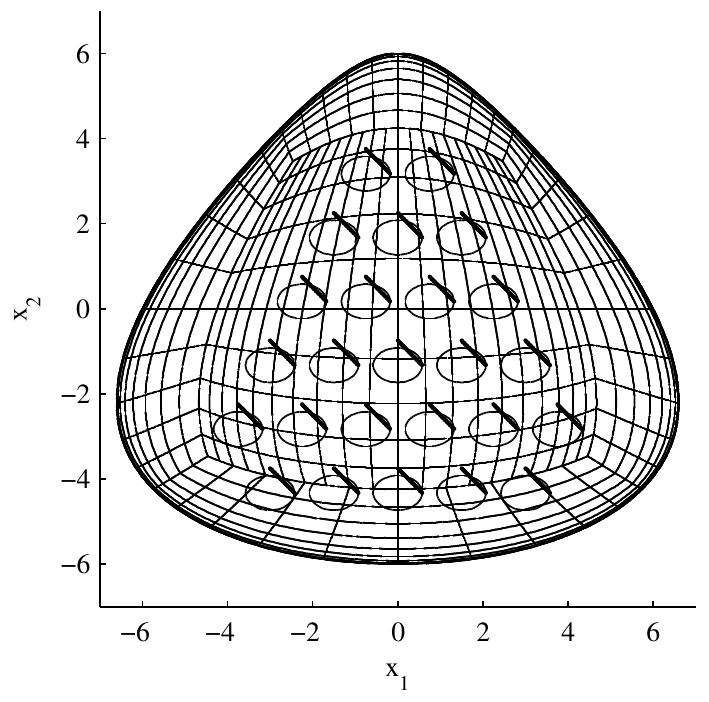} 
& 
\raisebox{3.2cm} 
{ 
$
\begin{array}{c} 
\includegraphics[scale=1]{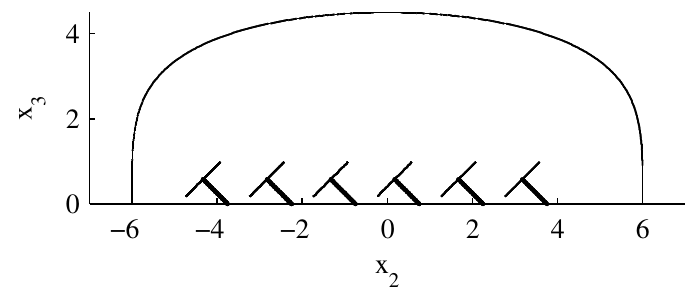} 
\\ \mbox{\lbc `Front' view, mesh outline only} \\ 
\includegraphics{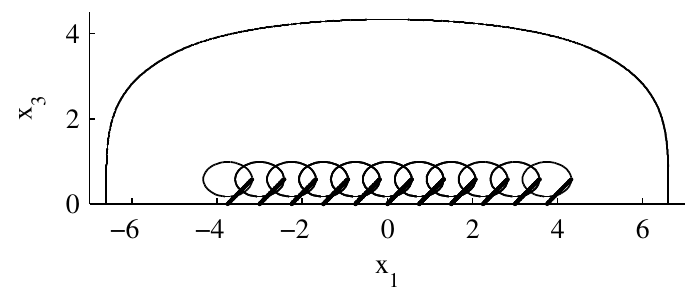} 
\end{array}  
$
}
\end{array}
$
\caption{The geometry of the node used for the computations. Ellipses show the trajectory of the cilium tip over its rotational beat; the rod-shaped cilia are shown in their `leftmost' position. For simulations in this paper, all cilia were synchronised, rotating in the clockwise direction viewed from above, and had the same geometric specification: tilt angle $\theta=35\degree$, semi-cone angle $\psi=45\degree$ and tilt direction $\phi=180\degree$, corresponding to a completely posterior tilt. (a) View from ventral to dorsal, looking onto the node from `above', with mesh. (b) View from `left' to `right', looking at the $x_2 x_3$ plane. (c) View from posteroir to anterior, looking at the $x_1 x_3$ plane.}
\label{fig:geometry}
\end{figure}

The upper membrane surface $(x_1,x_2,x_3)=\left(\hat{X},\hat{Y},\hat{Z}\right)$ was defined as a hemisphere, deformed using the following transformation:
\begin{eqnarray}
\hat{X} & = & X\left(1-\frac{Y}{2}\right) \mbox{,}\nonumber \\
\hat{Y} & = & Y \mbox{,} \nonumber \\
\hat{Z} & = & \frac{3}{4} Z^{1/2}.\label{mesh_deform}
\end{eqnarray}
where $(X,Y,Z)$ are the coordinates of a hemispherical surface $X^2+Y^2+Z^2=A^2$, $Z>0$, with radius $A=6$, as shown in
figure~\ref{fig:geometry}. The cilium beat cycle was defined by a tilted conical rotation, as in equation~\eqref{eq:conical}. All results were
calculated with tilt angle $\theta=35\degree$ and semi-cone angle $\psi=45\degree$, to correspond with the particle tracking study in \shortcite{Smith08}.

\subsection{Instantaneous flow with and without the overlying membrane}
\begin{figure}
$
\begin{array}{ll}
\mbox{\lba Apex of recovery stroke, with membrane} & \mbox{\lbb Apex of effective stroke, with membrane}\\ 
\includegraphics[scale=1]{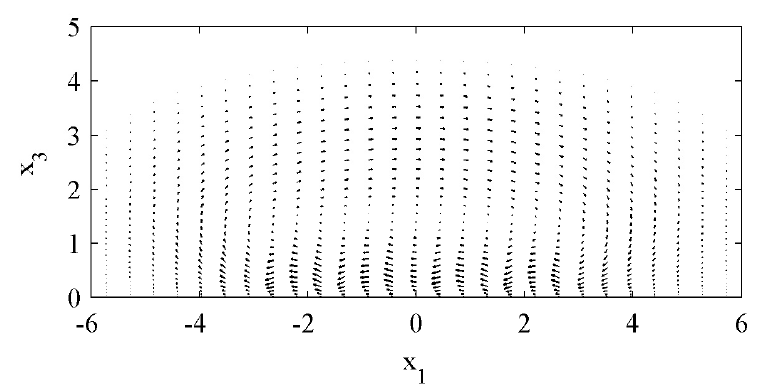} 
&
\includegraphics[scale=1]{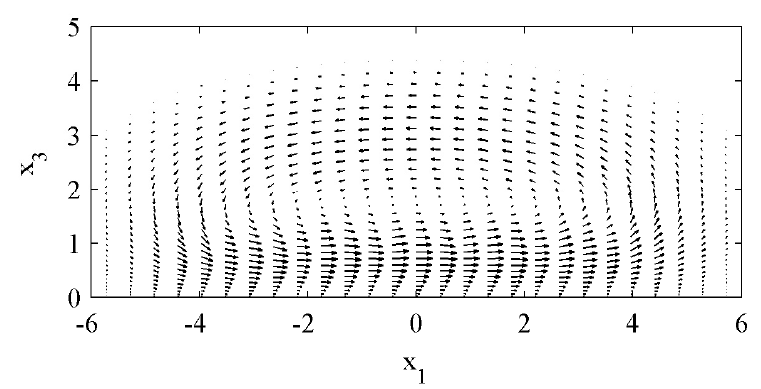}
\\
\mbox{\lbc Apex of recovery stroke, no membrane} 
&
\mbox{\lbd Apex of effective stroke, no membrane} \\
\includegraphics{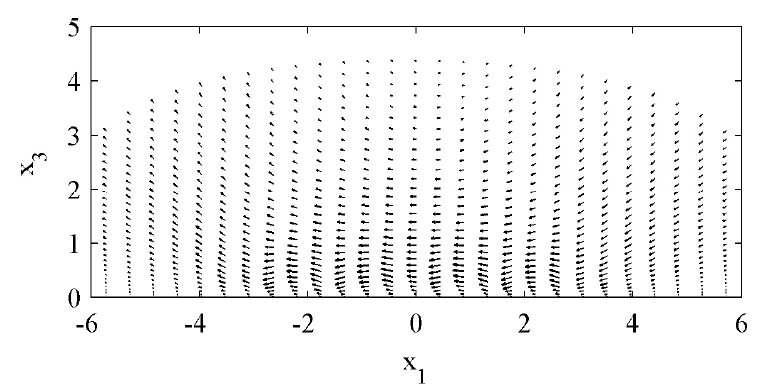} 
&
\includegraphics[scale=1]{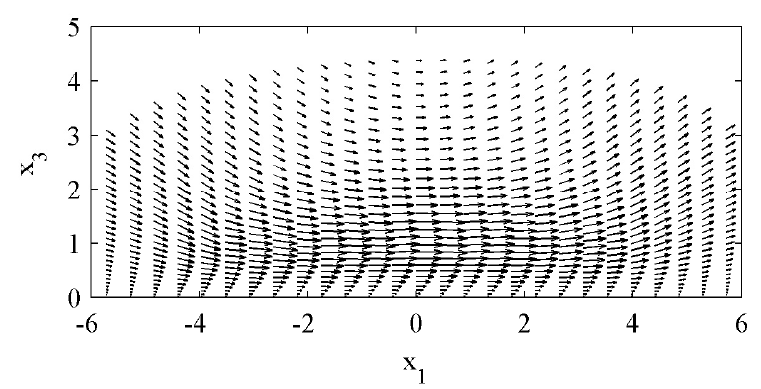}
\end{array}
$
\caption{Instantaneous flow field across the plane $x_2=-1.95$ at the apices of (a, c) the recovery stroke and (b, d) the effective stroke. Conventionally, the positive $x_1$ direction is referred to as the `leftward' direction, and appears as rightward in this figure.
(a, b) show results for simulations with the overlying Reichert's membrane boundary, (c, d) show results without the membrane boundary. To interpret the velocity scale, the peak velocity vector in figure (b) has dimensionless magnitude $0.63$, corresponding to a dimensional value of $19\;\mu$m/s for cilium length $3\;\mu$m and beat frequency $10$ Hz. By comparison, the peak velocity vector in figure (d) has magnitude $1.01$, or $30; \mu$m/s, showing that the upper membrane has a signficant effect even in the ciliated region.}
\label{fig:instantaneous}
\end{figure}

Figure~\ref{fig:instantaneous} shows instantaneous velocity field plots in the section $x_2=-1.95$, the nodal geometry being as shown in figure~\ref{fig:geometry}. The flow is shown at the apices of the recovery (a) and effective (b) strokes, that is at the highest and lowest points of the cilium beat cycle. 
The flow is in the direction of the cilia motion in the region $x_3<1.75$, there being a return flow in the region $x_3>1.75$ induced by the presence of the membrane. Due to the cilia being closer to the membrane during the recovery stroke, the induced flow magnitude is smaller, producing an overall positive `leftward' flow close to the cilia and a negative `rightward' flow in the upper region.
No instantaneous counter flow is evident close to the no-slip plane $x_3=0$. Figures~\ref{fig:instantaneous}(c, d) show the corresponding results across the same domain but with no overlying membrane. The principal difference is the complete absence of the upper return flow. Additionally, when the membrane is neglected, a `stresslet' type fluid flow is evident, with `upstream' fluid being drawn towards the epithelium
and `downstream' fluid being pushed away from the epithelium, as reported in earlier work \shortcite{Smith08}.

\subsection{Mean flow}
\begin{figure}
\begin{center}
$
\begin{array}{c}
\mbox{Isometric view} \\
\includegraphics{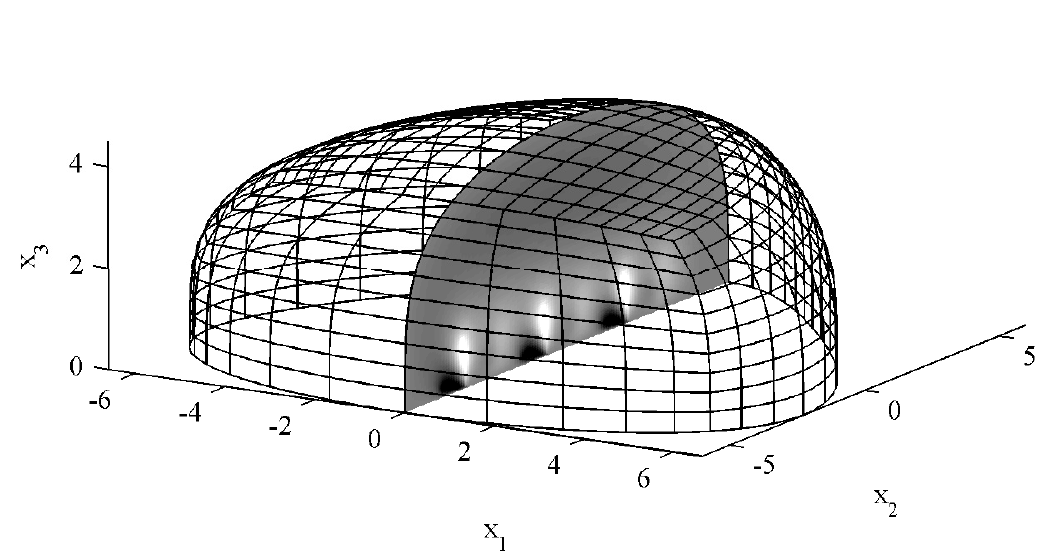} 
\\
\mbox{`Top' view} \\
\includegraphics{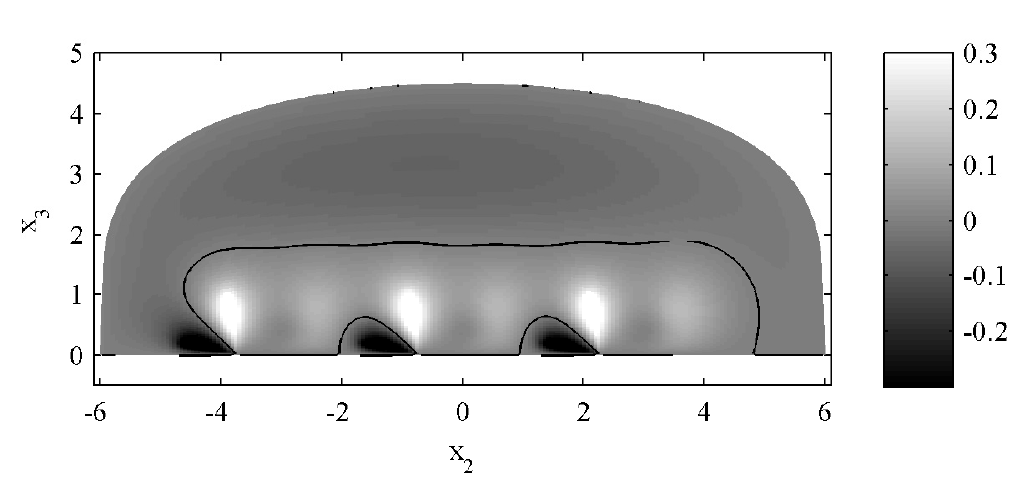}
\end{array}
$
\end{center}
\caption{Time-averaged `leftward' flow component $\bar{u}_1$, over the medial plane of the node $x_1=0$, shown in an isometric view, with mesh, and as a sagittal projection onto the $x_2 x_3$ plane. Black indicates negative transport, white positive transport, with contours showing the zero mean flow lines $\bar{u}_1=0$, which separate the positive and negative flow regions.}
\label{fig:sagittal}
\end{figure}

\begin{figure}
\begin{center}
$
\begin{array}{c}
\mbox{Isometric view} \\
\includegraphics{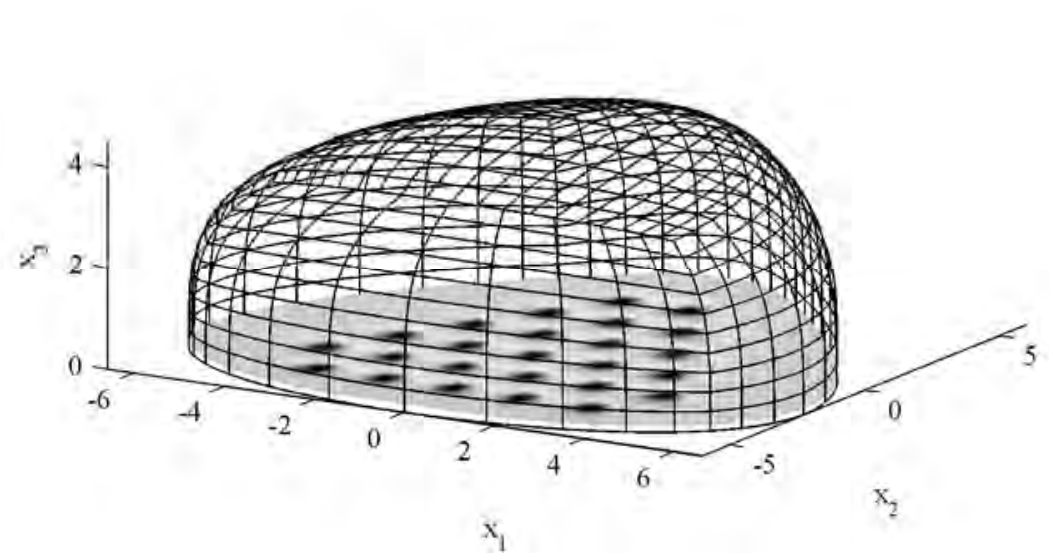}   
\\
\mbox{`Top' view} \\
\includegraphics{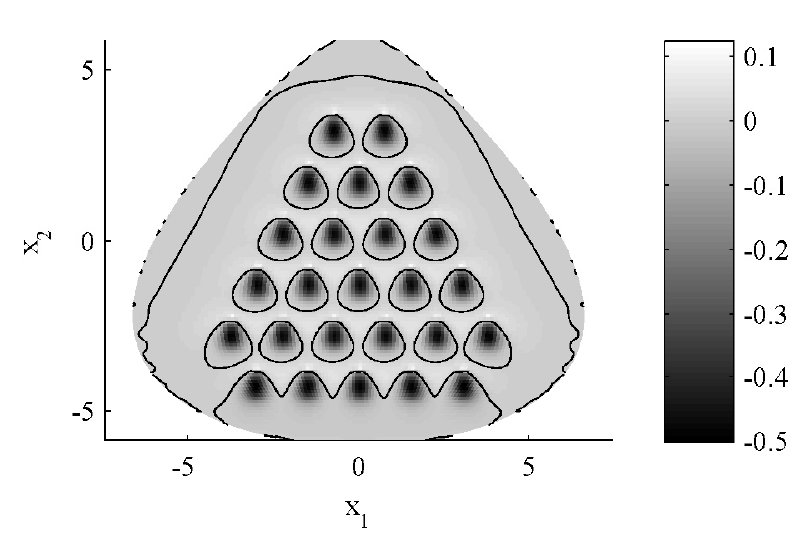} 
\end{array}
$
\end{center}
\caption{Time-averaged `leftward' flow component $\bar{u}_1$, over the plane $x_3=0.1125$, close to the epithelium. `Isometric' view, shows the membrane computational mesh; the `top' view shows the projection onto the $x_1 x_2$ plane, with contours showing the lines $\bar{u}_1=0$ separating regions of positive and negative mean flow.}
\label{fig:zslice}
\end{figure}

We now consider the time-averaged flow produced over an entire beat cycle; results were computed by averaging a beat cycle divided into $60$ discrete intervals and calculating the velocity component $u_1$ on the sagittal midplane $x_2=0$. The right-to-left component of the velocity field is shown in figure~\ref{fig:sagittal} as a shaded plot, the upper panel giving an in-situ view of the sagittal midplane, the lower panel showing a projection, with contours depicting the zero mean flow files separating leftward (positive) and rightward (negative) mean flow. The region of leftward flow is smaller in size than the region of rightward flow; additionally this section reveals `pockets' of mean rightward flow close to the epithelium, induced by the return stroke of the cilia. The sagittal section does not however reveal the extent of these regions; for this we examine the constant height $x_3=0.1125$ section shown in figure~\ref{fig:zslice}. This figure shows that throughout the majority of the node, the negative flow regions are localised to the cilia; between cilia there is a multiply-connected region of relatively slower positive flow. In contrast, the cilia close to the posterior edge of the node create a continuous region of negative flow; additionally there is a weak negative flow towards the left, right and anterior peripheries. The effect of this mean flow profile on particles within the node is unclear; we now examine particle tracking results to give insight into this.

\subsection{Particle Transport}
Figures~\ref{fig:tracking1}--\ref{fig:tracking5} show the results of particle tracking simulations, a simple model for the advection of finite-sized NVPs by the nodal flow. Results are shown for a period of $20000$ timesteps, or $333$ beat cycles, corresponding to approximately $30$ seconds at $10$ Hz. Figure~\ref{fig:tracking1} shows tracking results with initial particle position chosen for comparison with results given previously \shortcite{Smith08}. The particles are initially at the locations indicated by arrows; the lowest particle is at $x_3=0.1$, the highest at $x_3=1.1$.
All particles are initially swept to the left, even those starting in the region with negative mean flow (figure~\ref{fig:zslice}). This is because the rotational inner-field propels the particles to the region where the leftward flow dominates. Once transported to the left, all particles considered were subsequently lifted above the cilia array into the return flow region; indeed the particle starting closest to the epithelium was lifted the highest. Particles were generally not transported beyond the extent of the cilia array, and indeed particles starting higher up travelled a shorter distance before being returned. The eventual location of the particle varies considerably with a small change to initial position, as found previously \shortcite{Smith07} and observed experimentally \shortcite{Supatto08}.

The behaviour of particles released in the extremities of the node is shown in figures~\ref{fig:tracking2}--\ref{fig:tracking5}. Particles released to the left of the node (figure~\ref{fig:tracking2}) are generally lifted up into the return flow, although for particles released close to the epithelium, this process may not be complete in the time simulated (figure~\ref{fig:tracking2}(a)). The particle released at $x_3=1.1$ is returned sufficiently quickly that it re-enters the positive flow region, moving around the edges of two cilia envelopes. Figure~\ref{fig:tracking3} shows the transport of particles released just to the `right' of the cilia array. Particles are advected to the `left', although the trajectories and eventual  locations are highly unpredictable, with particles initially in a column, sometimes being temporarily trapped by one or two cilia vortices,
and particles being found on opposite sides of the node after 20000 timesteps. Figure~\ref{fig:tracking4} shows trajectories with initial location at the posterior edge of the node. Particles at all heights from $x_3=0.1$ to $x_3=1.1$ are found to be transported to the right, and are pushed down a short distance towards the epithelium. For initial height $0.5$ and above, the simulations predicted particles will orbit the cilium with base at $(-3.75,-2.25)$, before being carried to the left by a succession of cilia.

Figure~\ref{fig:tracking5} shows that particles released in the region anterior to the cilia array are subject to very little advection over a period of the duration of 333 beat cycles, at heights $x_3=0.1$ and $x_3=1.1$; results at other heights were similar. 

\subsection{Relation to experimental observations}
\begin{figure}
\begin{center}
$
\begin{array}{llll}
&\mbox{\lba `Top'} & \mbox{\lbb `Top'}&\\
&
\includegraphics[scale=1]{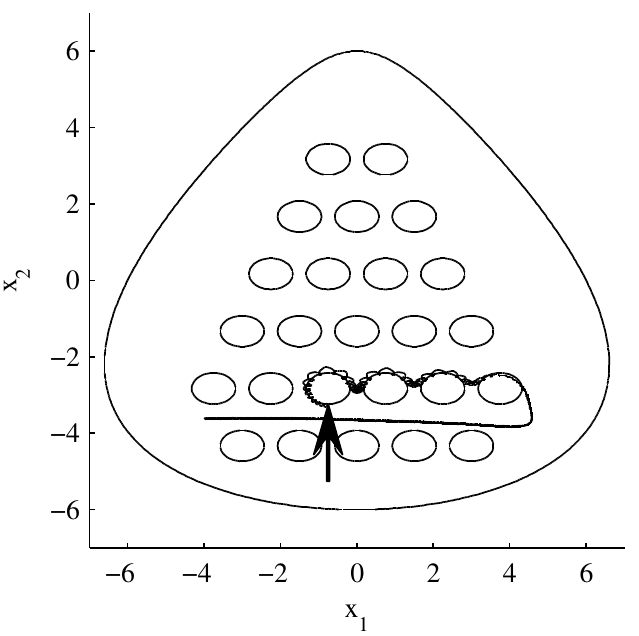}    
&
\includegraphics[scale=1]{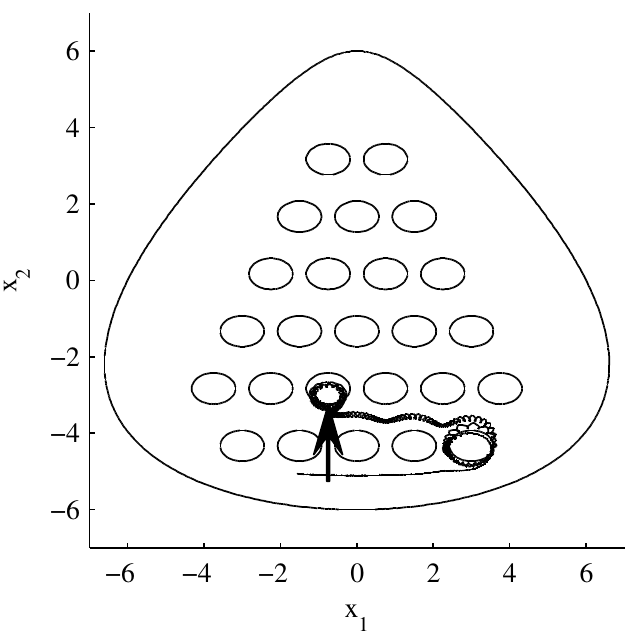}  
& \\
&\mbox{`Front'} & \mbox{`Front'}& \\
\begin{sideways}`RIGHT' \end{sideways}&
\includegraphics[scale=1]{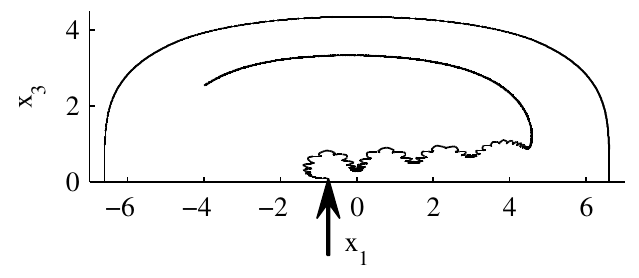}  
& 
\includegraphics[scale=1]{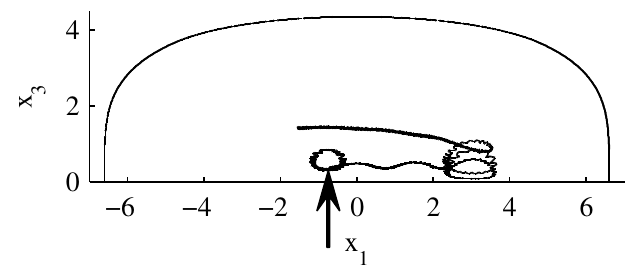}
&\begin{sideways}`LEFT'\end{sideways} \\
&\mbox{\lbc `Top'} & \mbox{\lbd `Top'}&\\
&
\includegraphics[scale=1]{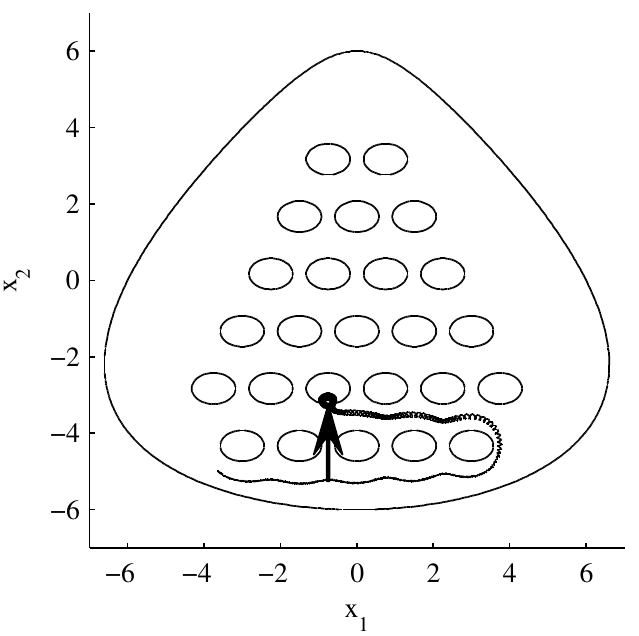}  
& 
\includegraphics[scale=1]{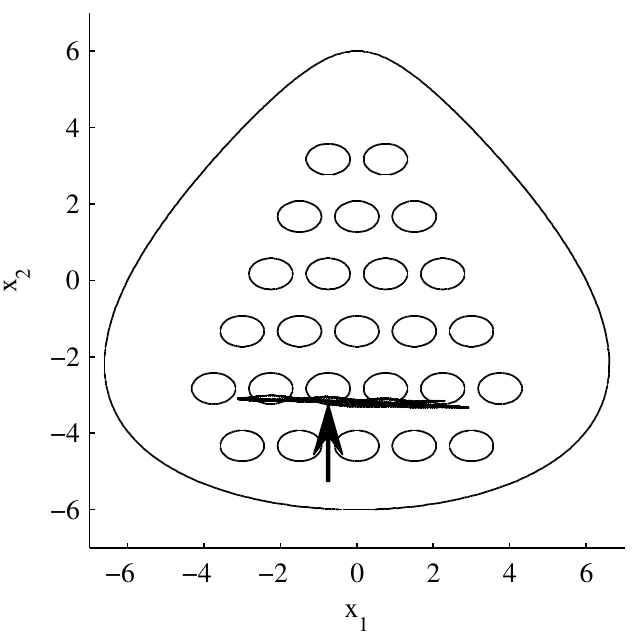}  
& \\
&\mbox{`Front'} & \mbox{`Front'}& \\
&
\includegraphics[scale=1]{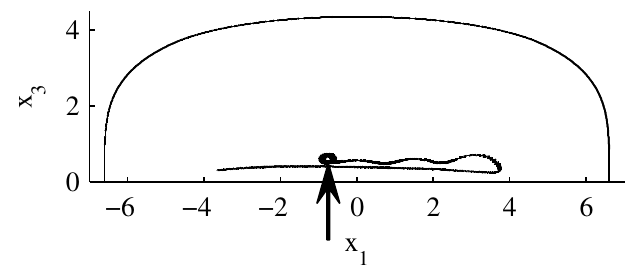}  
& 
\includegraphics[scale=1]{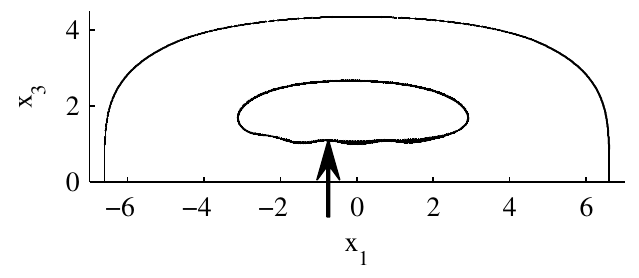}
&
\end{array}
$
\end{center}
\caption{Particle transport, initial particle position marked with arrow. The ellipses depict the envelope covered by the cilia tips viewed from above. (a) Initial position $x_1=-0.75$, $x_2=-3.25$, $x_3=0.1$, showing $x_1 x_2$ projection (`top') and $x_1 x_3$ projection (`front') . (b) Initial position as before, but with $x_3=0.3$. (c) Initial position as before but with $x_3=0.5$. (d) Initial position as before but with $x_3=1.1$.}
\label{fig:tracking1}
\end{figure}

\begin{figure}
\begin{center}
$
\begin{array}{llll}
&\mbox{\lba `Top'} & \mbox{\lbb `Top'}&\\
&
\includegraphics[scale=1]{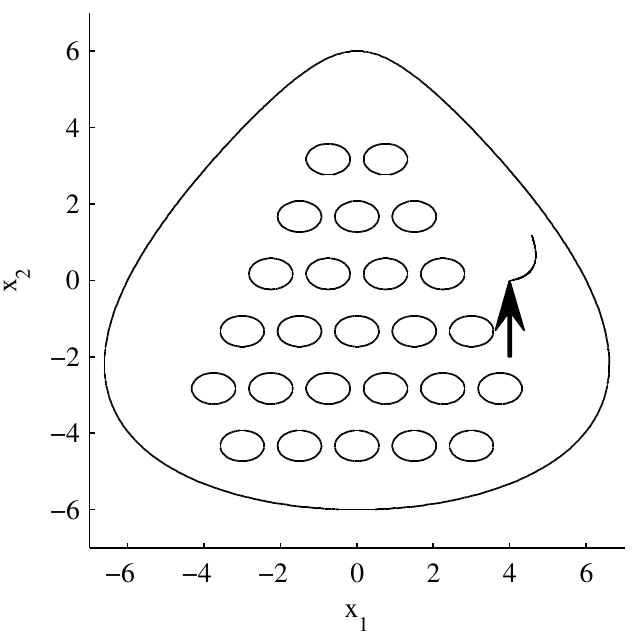}    
&
\includegraphics[scale=1]{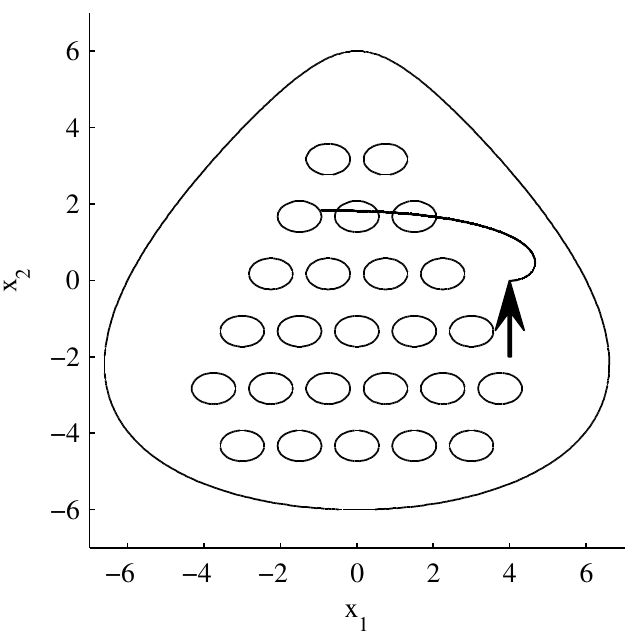}  
&\\
&\mbox{`Front'} & \mbox{`Front'} &\\
\begin{sideways}`RIGHT' \end{sideways}
&
\includegraphics[scale=1]{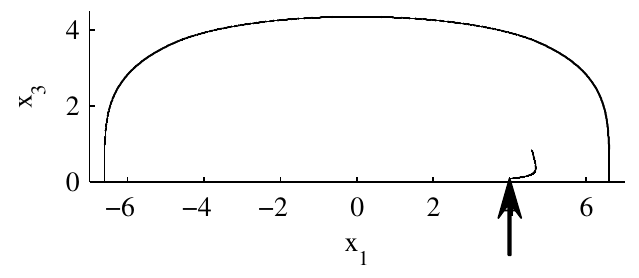} 
&
\includegraphics[scale=1]{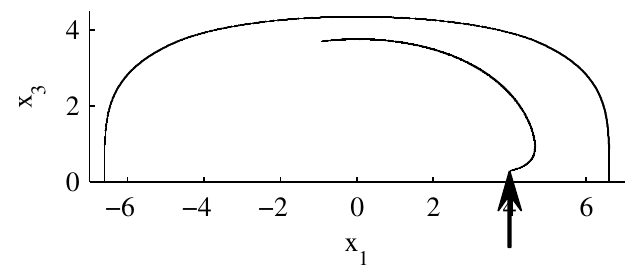} 
&
\begin{sideways}`LEFT'\end{sideways} \\
&\mbox{\lbc `Top'} & \mbox{\lbd `Top'}&\\
&
\includegraphics[scale=1]{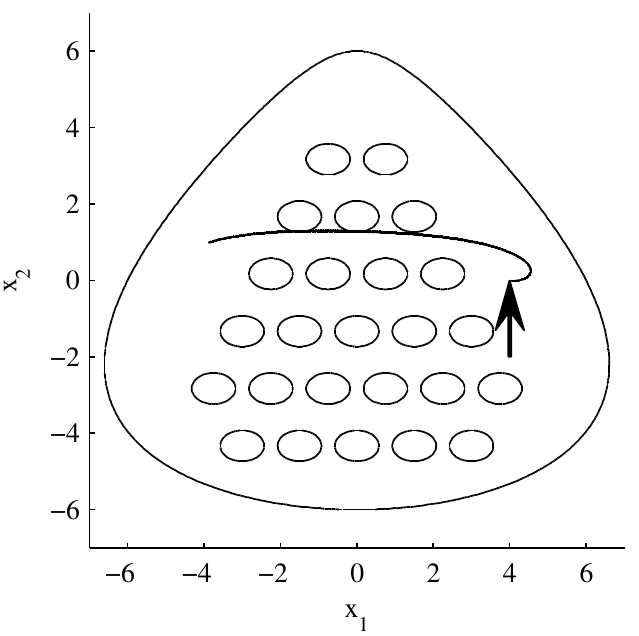}    
&
\includegraphics[scale=1]{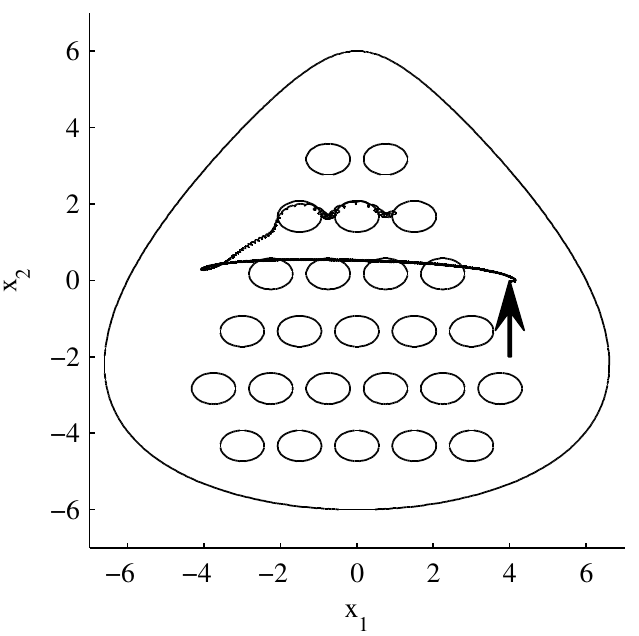}    
&
\\
&\mbox{`Front'} & \mbox{`Front'} &\\
&
\includegraphics[scale=1]{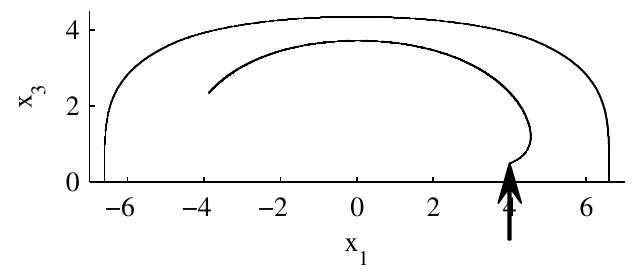} 
&
\includegraphics[scale=1]{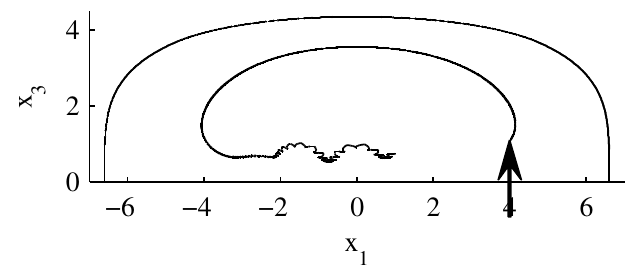}
&
\end{array}
$
\end{center}
\caption{Particle transport, initial particle position marked with arrow. (a) Initial position $x_1=4$, $x_2=0$, $x_3=0.1$, showing $x_1 x_2$ projection (`top') and $x_1 x_3$ projection (`front') . (b) Initial position as before, but with $x_3=0.3$. (c) Initial position as before but with $x_3=0.5$. (d) Initial position as before but with $x_3=1.1$.}
\label{fig:tracking2}
\end{figure}

\begin{figure}
\begin{center}
$
\begin{array}{llll}
&\mbox{\lba `Top'} & \mbox{\lbb `Top'}&\\
&
\includegraphics[scale=1]{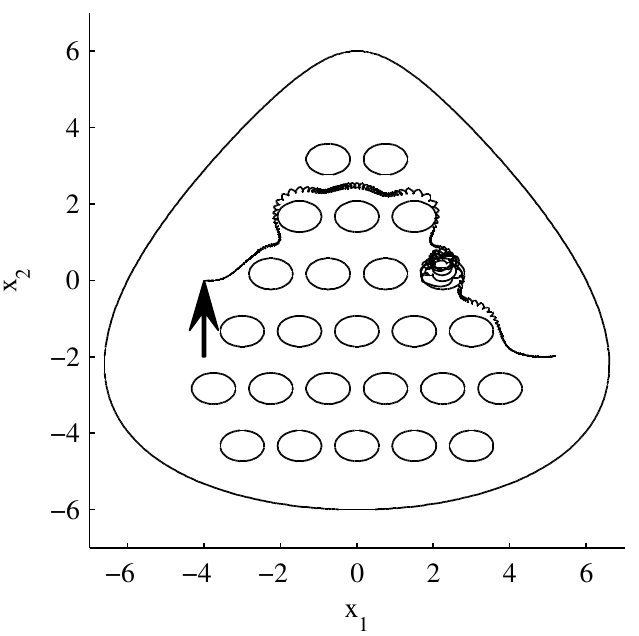}    
& 
\includegraphics[scale=1]{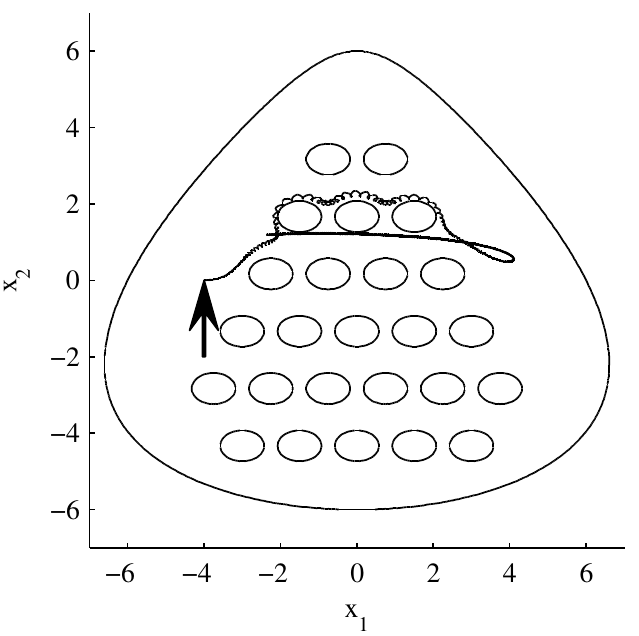}  
&\\
&\mbox{`Front'} & \mbox{`Front'} &\\
\begin{sideways}`RIGHT' \end{sideways}&
\includegraphics[scale=1]{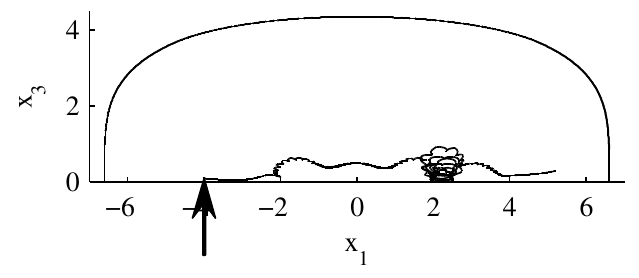} 
&  
\includegraphics[scale=1]{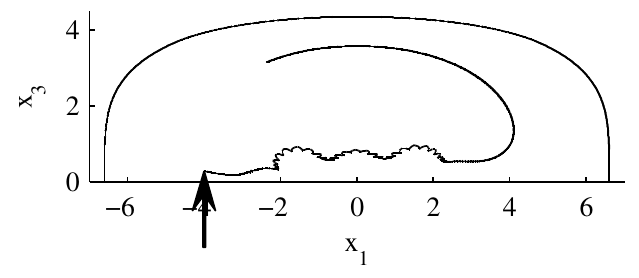}
&\begin{sideways}`LEFT'\end{sideways} \\
&\mbox{\lbc `Top'} & \mbox{\lbd `Top'}&\\
&
\includegraphics[scale=1]{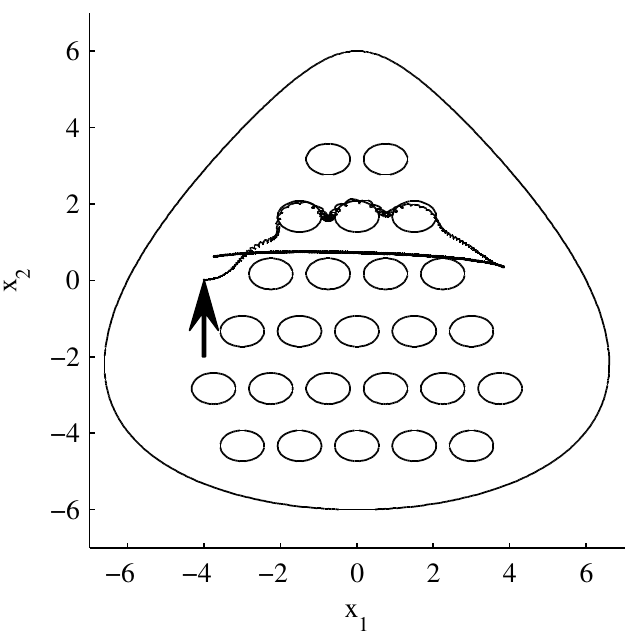}    
& 
\includegraphics[scale=1]{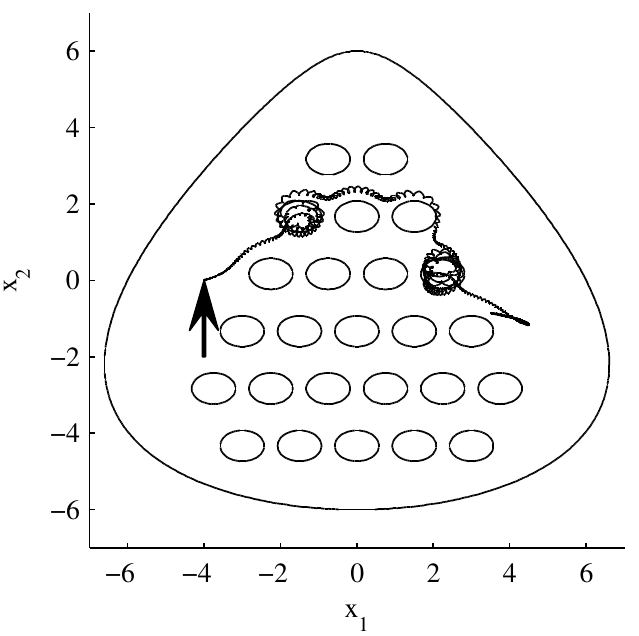}    
&\\
&\mbox{`Front'} & \mbox{`Front'} &\\
&
\includegraphics[scale=1]{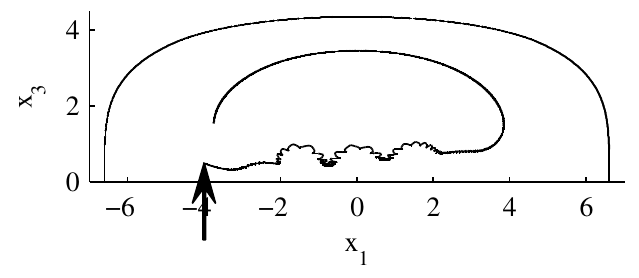} 
& 
\includegraphics[scale=1]{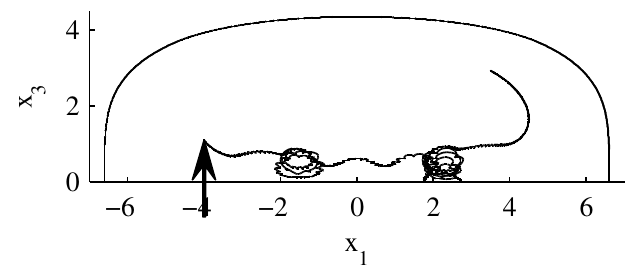}
&
\end{array}
$
\end{center}
\caption{Particle transport, initial particle position marked with arrow. (a) Initial position $x_1=-4$, $x_2=0$, $x_3=0.1$, showing $x_1 x_2$ projection (`top') and $x_1 x_3$ projection (`front') . (b) Initial position as before, but with $x_3=0.3$. (c) Initial position as before but with $x_3=0.5$. (d) Initial position as before but with $x_3=1.1$.}
\label{fig:tracking3}
\end{figure}

\begin{figure}
\begin{center}
$
\begin{array}{llll}
&\mbox{\lba `Top'} & \mbox{\lbb `Top'}&\\
&
\includegraphics[scale=1]{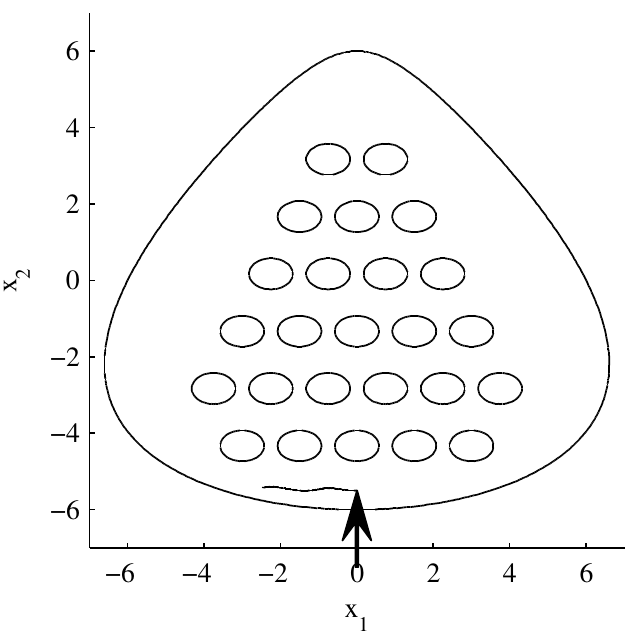}    
& 
\includegraphics[scale=1]{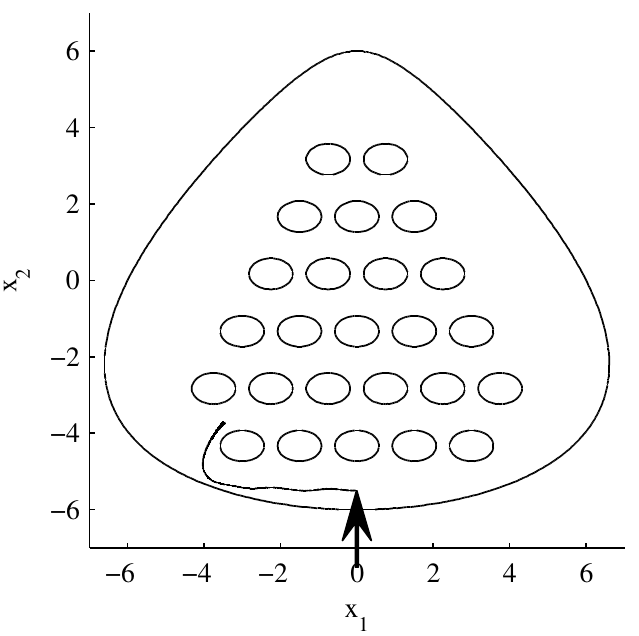}  
&
\\
&\mbox{`Front'} & \mbox{`Front'} &\\
\begin{sideways}`RIGHT' \end{sideways}&
\includegraphics[scale=1]{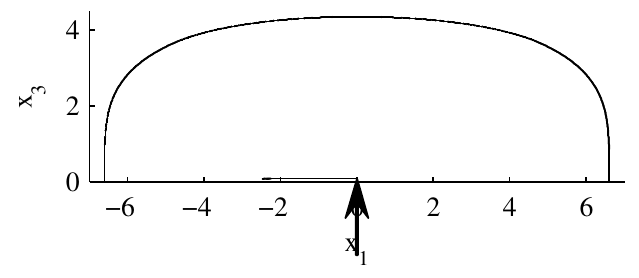} 
& 
\includegraphics[scale=1]{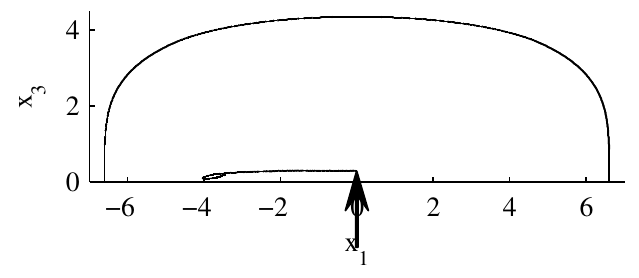} 
&
\begin{sideways}`LEFT'\end{sideways} \\
&\mbox{\lbc `Top'} & \mbox{\lbd `Top'}&\\
&
\includegraphics[scale=1]{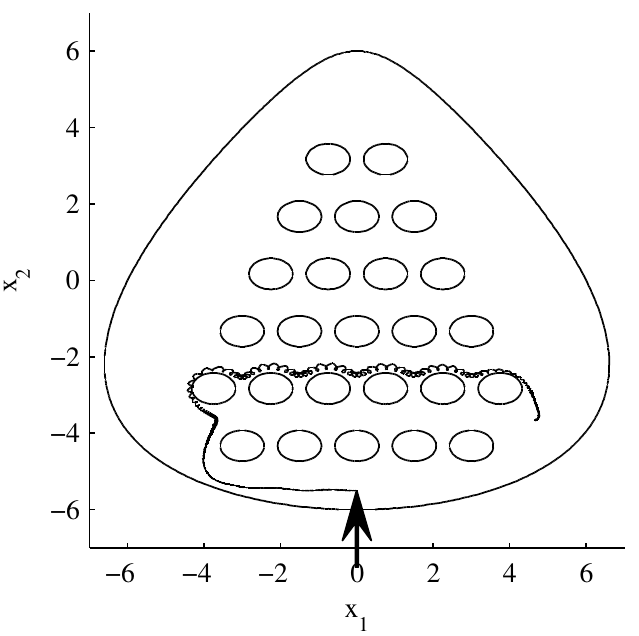}    
& 
\includegraphics[scale=1]{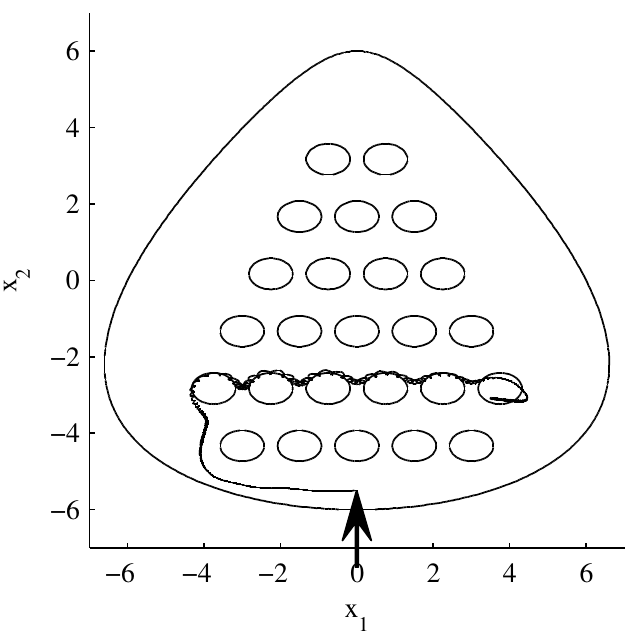}    
&
\\
&\mbox{`Front'} & \mbox{`Front'} &\\
&
\includegraphics[scale=1]{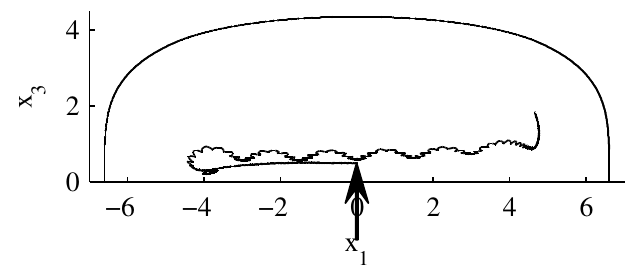} 
& 
\includegraphics[scale=1]{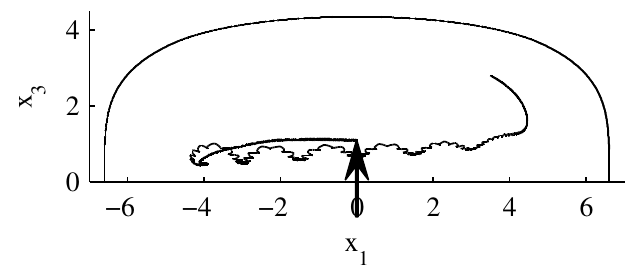}
&
\end{array}
$
\end{center}
\caption{Particle transport, initial particle position marked with arrow. (a) Initial position $x_1=0$, $x_2=-5.5$, $x_3=0.1$, showing $x_1 x_2$ projection (`top') and $x_1 x_3$ projection (`front') . (b) Initial position as before, but with $x_3=0.3$. (c) Initial position as before but with $x_3=0.5$. (d) Initial position as before but with $x_3=1.1$.}
\label{fig:tracking4}
\end{figure}

\begin{figure}
\begin{center}
$
\begin{array}{llll}
&\mbox{\lba `Top'} & \mbox{\lbb `Top'}&\\
\begin{sideways}`RIGHT'\end{sideways} &
\includegraphics[scale=1]{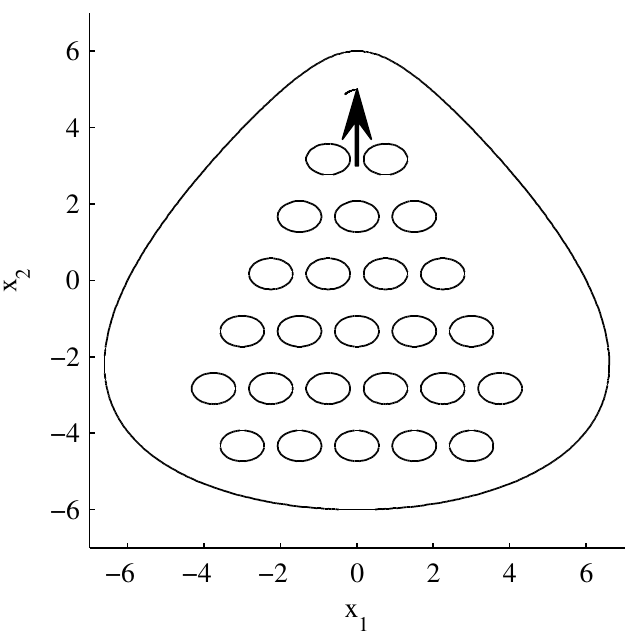}    
& 
\includegraphics[scale=1]{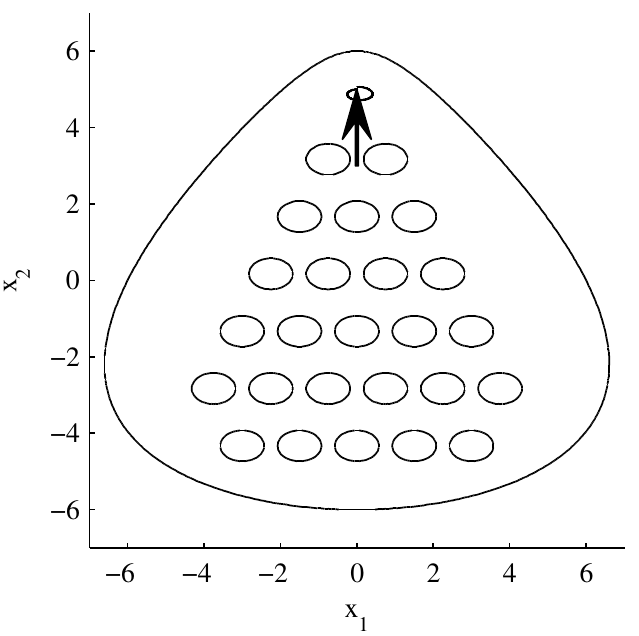}  
& \begin{sideways}`LEFT'\end{sideways} \\
&\mbox{`Front'} & \mbox{`Front'} & \\
& 
\includegraphics[scale=1]{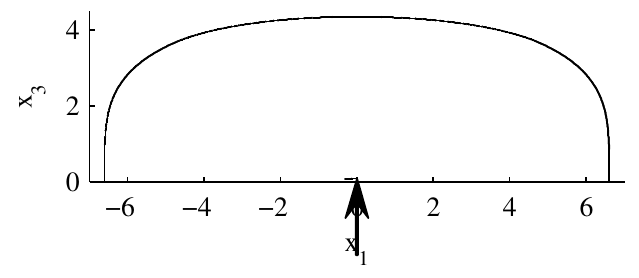} 
& 
\includegraphics[scale=1]{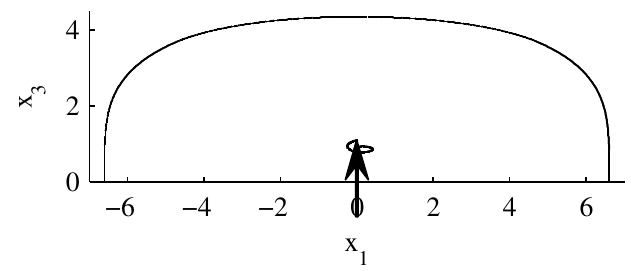} 
&   
\end{array}
$
\end{center}
\caption{Particle transport, initial particle position marked with arrow. (a) Initial position $x_1=0$, $x_2=5$, $x_3=0.1$, showing $x_1 x_2$ projection (`top') and $x_1 x_3$ projection (`front') . (b) Initial position as before, but with $x_3=1.1$}
\label{fig:tracking5}
\end{figure}

The instantaneous `mainstream' flow velocity in regions between the cilia, as shown in figure~\ref{fig:instantaneous} typically has a peak value of around $0.5$, corresponding to $15$--$25\; \mu$m/s, while the particle drift speed can be two orders of magnitude slower, with the leftward drift phase in figure~\ref{fig:tracking1}(a) having a dimensionless speed of approximately $0.09$, in dimensional units this corresponds to $2.7$--$4.5\;\mu$m/s, the rightward drift phase being slower still, the dimensionless speed being around $-0.036$, corresponding to $-1.1$--$-1.8\;\mu$m/s.
By comparison, Okada et al. \shortcite{Okada05} reported positive flow drift speeds with a mean of around $3.5\;\mu$m/s in the mouse and return flow drift speeds of around $-1.5\;\mu$m/s; in both cases there was considerable range in the observations, with for example some particles having a drift speed of up to $6\;\mu$m/s.

\section{Summary and discussion: research to date and new findings}
Previous models have established a number of important features of the generation of the nodal flow:
\begin{itemize}
\item The flow is generated by whirling cilia tilted towards the posterior, approximately tracing out the surface of an inclined cone \shortcite{Cartwright04,Nonaka05,Okada05}. 
\item Fluid dynamic models predict optimal fluid flow when the axis of the cone is tilted by approximately $35\degree$ and the semi-cone angle is approximately $55\degree$, in general agreement with an experimental analogue \shortcite{Nonaka05} and biological observations \shortcite{Nonaka05,Okada05,Supatto08}.
\item The viscous drag produced by the cell surface results in the relatively upright effective stroke, producing a much larger `near-field' where the velocity is ${O}(1/r)$ than the recovery stroke. Moreover, the strength of the ${O}(1/r^2)$ far-field velocity is greater during the effective stroke than the recovery stroke, being proportional to the height of the cilium above the surface. The combined effect is for more fluid to be transported to the left by the effective stroke than is returned to the right by the recovery stroke, resulting in an overall leftward flow as summarised in the schematic figure~\ref{fig:generation_nodal_flow}.
\item The resulting particle paths are a `loopy drift' close to the cilia, and a `radial drift' further away. 
\item From a time-averaged perspective, image singularities convert a near-field rotational motion, characterised by an antisymmetric rotlet tensor, into a far-field straining motion, characterised by a symmetric stresslet tensor. As particles are transported by the array of cilia they may move from the far-field of one cilium to the near-field of another and vice versa (`multiple-cilia effect').
\item These effects can be understood, and modelled mathematically, through the use of image singularities located beneath a plane boundary, as first derived by Blake \shortcite{Blake71a}. 
\item The multiple-cilia effect may create chaotic advection, as observed experimentally \shortcite{Supatto08}.
\item The closure of the nodal cavity will create a return flow in the upper region of the node \shortcite{Cartwright07}.
\end{itemize}
This study has developed the model further by combining time-dependent cilia-driven flow with a closed nodal cavity. It was found that:
\begin{itemize}
\item The region of negative flow is larger than the region of positive flow, covering most of the region above the cilia tips, and in addition for the model considered the extreme peripheries of the node beyond the cilia array.
\item The presence of the membrane significantly reduces the flow rate in the ciliated region. 
\item Close to the epithelium, there is generally a slow positive mean flow, with a negative mean flow in enclosed regions around the cilia. There may also be a layer of continuous negative flow close to the epithelium at the posterior end of the node.
\item Particles released between the epithelium and the height of the cilia tips are generally advected to the left edge of the ciliated array, before being lifted into the return flow region. The trajectories taken may however be very unpredictable, involving detours around cilia vortices for one or many rotations, and considerable vertical mixing of the flow. 
\item Particles released in the posterior negative flow layer are transported to the right of the node, before entering a leftward trajectory. This behaviour is not present in models which neglect the upper membrane. 
\item Particles released at the anterior extremity of the node were not predicted to travel a significant distance over a simulated period of 330 beat cycles, corresponding to about 30 seconds. In this region especially, diffusion may be the dominant transport mechanism.
\end{itemize}

Our model does not yet explain how NVPs may be trapped and/or broken up by cilia at the left of the node: once the particles have reached the edge of the cilia array they typically are transferred to the return flow. It may be that cilia must extend to the edge of the node in order for this to occur, and perhaps that particle break-up and endocytosis \shortcite{Yu09} or some other form of morphogen inactivation \shortcite{Cartwright04} must prevent morphogens being returned to the right and hence spread equally.

Chaotic low Reynolds number advection has been predicted to assist with the feeding of sessile micro-organisms on the basis of a stokeslet-image system model, using similar techniques to those described in this paper \shortcite{Blake96}. 
While we have reported qualitatively `chaotic' behaviour, it remains to quantify mathematically the degree and extent of chaos in nodal flow, for example through Lyapunov exponents \shortcite{Blake96}. 
The apparent presence of chaotic transport is perhaps surprising in a structure that has evolved to break symmetry in an organised way. It is unclear to the present authors to what extent chaos may help, hinder, or have no significant effect on the symmetry-breaking process, and how the biological system may have evolved either to minimise chaos, or to exploit its potential to spread particles across the left side of the node.

\section{Future directions}
The ventral node structure and equivalent organising structures vary considerably in shape, size and other detailed aspects across different classes of vertebrates. Fluid mechanical modelling has a role to play in determining the importance of different aspects of node morphology and scale.
The biological system differs from our idealised mathematical model in a number of respects, not least the heterogeneity in cilia beat pattern, cilia orientation and sometimes direction, in addition to irregular spacing, as shown in biological data. The epithelium itself is also not a smooth planar surface, but rather is made up of a `cobbled' surface of curved cells. The regularized stokeslet method described in this study provides a basis for models which could take this complex geometry into account. 
The development and growth of the node and expression of cilia is also a dynamic process, and the detailed effects of these dynamic changes is yet to be uncovered. An additional important feature not considered in this study is the effect of synchronised cilia, versus cilia beating with different phase, or indeed different frequencies. Results reported previously \shortcite{Smith07} suggest that this may significantly alter the trajectories of particles released near the cilia envelopes, although it does not significantly change the overall particle trajectories. Further investigation of this effect within an enclosed nodal cavity is warranted.

The end-point of these models is the determination of how the flow may establish a morphogen gradient, and how this will induce asymmetric signalling. Initial mathematical and laboratory models of this phenomenon have been developed by a number of groups \shortcite{Cartwright07,Cartwright04,Okada05,Supatto08}, leading to a number of important insights. In 1970, Crick \shortcite{Crick70} proposed that diffusion through cells, combined with a `source-sink' process, could produce spatial gradients of morphogens in embryogenesis. Recently, Yu et al. \shortcite{Yu09} demonstrated  that a morphogen protein, Fgf8, can form gradients by diffusing into the extracellular fluid, the `sink' function being performed by the endocytosis. These experiments were performed on living zebrafish embryos during gastrulation, the developmental stage at which left-right symmetry breaking is initiated. The diffusion measurements were achieved through the use of fluorescence correlation spectroscopy; future studies with novel techniques such as this will allow accurate quantification of morphogen gradients in the node associated with left-right symmetry breaking. Fluid mechanics plays an important role in many aspects of both animal and plant developmental biology, as recently reviewed by Cartwright and co-authors \shortcite{Cartwright09}, and presents many unexplored problems at the interface of mathematical modelling and experimental biology. 

The past decade has seen significant advances in our biological understanding of the early stages of this process. These advances have inspired fluid dynamics studies, and on certain occasions have been anticipated by them. We look forward to future multidisciplinary work on this exciting and fundamental scientific problem.

\section*{Acknowledgements}
This paper represents the influence Professor Ernie Tuck had on the last author and in turn his academic grandchildren. 
The authors thank Dr Eamonn Gaffney, Mathematical Institute, University of Oxford for his contribution to techniques and models used in earlier work that formed the basis for this study. The authors also thank three anonymous referees for their constructive comments. AAS acknowledges an Engineering and Physical Sciences Research Council doctoral training award; part of this research was developed while DJS was funded by a Medical Research Council Training Fellowship in Computational Biology (G0600178), which also provided support to JRB.


\end{document}